\documentclass[12pt]{article}
\usepackage{epsfig}\parskip 5pt plus 1pt
\newcommand{\nn}{\nonumber}

\newcommand{\be}{\begin{equation}}
\newcommand{\ee}{\end{equation}}
\newcommand{\bea}{\begin{eqnarray}}
\newcommand{\eea}{\end{eqnarray}}

\newcommand{\dd}{\displaystyle}

\begin{document}
%
%%%%%%%%%%%%%%%%%%%%%%%%%%%%%%%%%%%%%%%%%%%%%%%%%%%%%%%%%%%
%
\thispagestyle{empty}
\begin{flushright}

{CERN-TH/2003-297} \\
{IFT-UAM/CSIC-03-51} \\
{UG-FT-160/03} \\
{CAFPE-30/03} \\
{\tt hep-ph/0312072}

\end{flushright}
\vspace*{1cm}
%
%%%%%%%%%%%%%%%%%%%%%%%%%%%%%%%%%%%%%%%%%%%%%%%%%%%%%%%%%%%%%%%%
\begin{center}
%%%%%%%%%%%%%%%%%%%%%%%%%%%%%%%%%%%%%%%%%%%%%%%%%%%%%%%%%%%%%%%%
{\Large{\bf Clone flow analysis for a theory inspired Neutrino 
Experiment planning} }\\
\vspace{1.5cm}
A. Donini$^{\rm a,}$\footnote{E-mail: andrea.donini@roma1.infn.it},
D. Meloni$^{\rm b,}$\footnote{E-mail: davide.meloni@roma1.infn.it} and
S. Rigolin$^{\rm a c,}$\footnote{E-mail: stefano.rigolin@cern.ch}\\
\vspace{.5cm}
{\small $^{\rm a}$ Dep. F\'{\i}sica Te\'orica, Universidad Autonoma Madrid, 
E-28049  Madrid, Spain \\
$^{\rm b}$ Dep. de F\'{\i}sica Te\'orica y del Cosmos, Universidad de Granada, \\
Campus de Fuente Nueva, E-18002 Granada, Spain \\
$^{\rm c}$ Theory Division, CERN, 1211 Geneve, Switzerland}
%%%%%%%%%%%%%%%%%%%%%%%%%%%%%%%%%%%%%%%%%%%%%%%%%%%%%%%%%%%%%%%%
\end{center}
%%%%%%%%%%%%%%%%%%%%%%%%%%%%%%%%%%%%%%%%%%%%%%%%%%%%%%%%%%%%%%%%
\begin{abstract}
The presence of several clone solutions in the simultaneous measurement 
of ($\theta_{13},\delta$) has been widely discussed in literature. In 
this letter we write the analytical formul\ae~of the clones location in the 
($\theta_{13},\delta$) plane as a function of the physical input pair 
($\bar\theta_{13},\bar\delta$). We show how the clones move with changing 
$\bar\theta_{13}$. The ``clone flow'' can be significantly different if 
computed (naively) from the oscillation probabilities or (exactly) from the 
probabilities integrated over the neutrino flux and cross-section.

%Using our complete computation we compare the clone flow of a set of possible 
%future neutrino experiments with different neutrino source: the CERN SuperBeam, 
Using our complete computation we compare the clone flow of a set of possible 
future neutrino experiments: the CERN SuperBeam, BetaBeam and Neutrino Factory 
proposals. We show that the combination of these specific BetaBeam and SuperBeam 
does not help in solving the degeneracies. On the contrary, the combination of 
one of them with the Neutrino Factory Golden and Silver channel can be used, 
from a theoretical point of view, to solve completely the eightfold degeneracy. 
\end{abstract}

\newpage
%%%%%%%%%%%%%%%%%%%%%%%%%%%%%%%%%%%%%%%%%%%%%%%%%%%%%%%%%%%

\section{Introduction}
\label{sec:theo}

The atmospheric and solar sector of the PMNS leptonic mixing matrix \cite{Pontecorvo:1957yb}
have been measured with quite good resolution by SK \cite{yanagisawa}, SNO \cite{Ahmed:2003kj}
and KamLand \cite{KAMLAND}. These experiments measure two angles, $\theta_{12}$ and $\theta_{23}$,
and two mass differences, $\Delta m^2_{12}$ and $\Delta m^2_{23}$ (for the explicit 
form of the PMNS matrix and the adopted conventions, see for example \cite{Donini:1999jc}).
The present bound on $\theta_{13}$, $\sin^2\theta_{13} \leq 0.02$, is extracted from the 
negative results of CHOOZ~\cite{chooz} and from three-family analysis of atmospheric and 
solar data. The PMNS phase $\delta$ is totally unbounded as no experiment is sensitive to 
the leptonic CP violation. The main goal of next neutrino experiments will be to measure 
these two still unknown parameters of the leptonic mixing matrix. In this paper we concentrate 
on analyzing, form a theoretical point of view, a particular problem that arise when trying 
to measure simultaneously ($\theta_{13},\delta$).

In \cite{Burguet-Castell:2001ez} it has been noticed that the appearance probability 
$P_{\alpha \beta} (\bar \theta_{13},\bar \delta)$ obtained for neutrinos at a fixed energy 
and baseline with input parameter ($\bar\theta_{13}, \bar\delta$) has no unique solution. 
Indeed, the equation:
\be
\label{eq:equi0} 
P_{\alpha \beta} (\bar \theta_{13}, \bar \delta)  =  P_{\alpha \beta} 
(\theta_{13}, \delta)
\ee 
has a continue number of solutions. The locus of ($\theta_{13},\delta$) satisfying this equation 
is called ``equiprobability curve''. Considering the equiprobability curves for neutrinos and 
antineutrinos with the same energy (and the same input parameters), the system of equations
\be
\label{eq:equi1} 
P^\pm_{\alpha \beta} (\bar \theta_{13},\bar \delta) = P^\pm_{\alpha \beta} 
(\theta_{13}, \delta)
\ee 
(where $\pm$ refers to neutrinos and antineutrinos) has two intersections: the input pair 
($\bar \theta_{13},\bar \delta$) and a second, energy dependent, point. This second intersection 
introduces an ambiguity in the measurement of the physical values of $\theta_{13}$ and $\delta$: 
the so-called {\it intrinsic clone}. As it was made clear 
in \cite{Minakata:2001qm,Barger:2001yr}, two other sources of ambiguities are present: 
\begin{itemize}
\item We only know the absolute value \cite{yanagisawa}, $|\Delta m^2_{23}|\in[1.9,3.5] \times 
     10^{-3}$ eV$^2$ of the atmospheric mass difference, not its sign. This ambiguity originates
     from the fact that atmospheric neutrino experiments 
     measure only the leading oscillation $\nu_\mu \to \nu_\tau$ that depends quadratically on 
     $\Delta m^2_{23}$;

\item We only know the allowed departure of $\theta_{23}$ from maximal mixing, 
      $\sin^2 2 \theta_{23} > 0.9$, but not if $\theta_{23}$ is actually smaller or greater 
      than $45^\circ$. This ambiguity originates in that neutrino experiments are looking for 
      $\nu_\mu,\nu_e$ disappearance or $\nu_\mu \to \nu_\tau$ oscillation, where $\theta_{23}$ 
      appears only through $\sin^2 2 \theta_{23}$.
\end{itemize}

As a consequence, future experiments will have as ultimate goal the measure of the two continuous 
variables $\theta_{13}$ and $\delta$ plus the two discrete variables: 
\bea
\bar s_{atm} &=& sign [ \Delta m^2_{23} ] \, , \\
\bar s_{oct} &=& sign [ \tan (2 \theta_{23}) ] \, .
\eea
These two discrete variables assume the values $\pm 1$, depending on the physical assignments of the 
$\Delta m^2_{23}$ sign ($s_{atm}=1$ for $m_3^2>m_2^2$ and $s_{atm}=-1$ for $m_3^2<m_2^2$) and of the 
$\theta_{23}$-octant ($s_{oct}=1$ for $\theta_{23}<\pi/4$ and $s_{oct}=-1$ for $\theta_{23}>\pi/4$).
As a consequence, taking into account all the present ignorance on the neutrino mixing and masses, 
eq.~(\ref{eq:equi1}) must be rewritten, more precisely, as:
\bea
\label{eq:equi0int} 
P^\pm_{\alpha \beta} (\bar \theta_{13},\bar \delta; \bar s_{atm},\bar s_{oct}) &=&
P^\pm_{\alpha \beta} (\theta_{13},\delta; s_{atm}=\bar s_{atm}; s_{oct}=\bar s_{oct})\, ,
\eea
where $\bar s_{atm}$ and $\bar s_{oct}$ have been included as input parameters in addition to 
$\bar \theta_{13}$ and $\bar \delta$. In eq.~(\ref{eq:equi0int}) we have implicitly assumed to 
know the right sign and the right octant for the atmospheric mass difference and angle. As these 
quantities are unknown (and presumably they will remain so at the time of the next neutrino 
facilities), all the following equiprobabilities systems of equations should be considered as well: 
\bea
\label{eq:equi0sign} 
P^\pm_{\alpha \beta} (\bar \theta_{13}, \bar \delta; \bar s_{atm}, \bar s_{oct}) &=&
P^\pm_{\alpha \beta} (\theta_{13}, \delta; s_{atm} = - \bar s_{atm}; s_{oct} = \bar s_{oct}) \, ,\\ 
\label{eq:equi0t23} 
P^\pm_{\alpha \beta} (\bar \theta_{13}, \bar \delta;  \bar s_{atm}, \bar s_{oct}) &=&
P^\pm_{\alpha \beta} (\theta_{13}, \delta; s_{atm} =  \bar s_{atm}; s_{oct} = -\bar s_{oct} )\, , \\
\label{eq:equi0t23sign} 
P^\pm_{\alpha \beta} (\bar \theta_{13}, \bar \delta;\bar s_{atm}, \bar s_{oct} ) &=&
P^\pm_{\alpha \beta} (\theta_{13}, \delta; s_{atm} = -  \bar s_{atm}; s_{oct} = - \bar s_{oct} ) \, .
\eea
These new sets of equiprobability systems arise when we equate the true probability (left hand side) 
with the probabilities obtained including one of the three possible wrong guesses on $s_{atm}$ and 
$s_{oct}$ (right hand side).

Solving the four systems of eq.~(\ref{eq:equi0int})-(\ref{eq:equi0t23sign}) will result 
in obtaining the true solution plus the appearance of additional {\it clones} to form an 
eightfold-degeneracy \cite{Barger:2001yr}. These eight solutions are respectively: 
\begin{itemize}
\item the true solution and its {\em intrinsic clone}, obtained solving the system 
      of eq.~(\ref{eq:equi0int});
\item the $\Delta m^2_{23}$-sign clones (hereafter called {\em sign clones}) of the 
      true and intrinsic solution, obtained solving the system of eq.~(\ref{eq:equi0sign});
\item the $\theta_{23}$-octant clones (hereafter called {\em octant clones}) of the 
      true and intrinsic solution, obtained solving the system of eq.~(\ref{eq:equi0t23});
\item the $\Delta m^2_{atm}$-sign $\theta_{23}$-octant clones (hereafter called 
      {\em mixed clones}) of the true and intrinsic solution, obtained solving the system 
      of eq.~(\ref{eq:equi0t23sign}).
\end{itemize}

In the following section we illustrate the technique used for calculating the location of the 
clone solutions. We describe explicitly the formul\ae~for the $\nu_e\to\nu_\mu$ oscillation 
probability, due to the great importance it has in literature. In fact this ``golden'' channel 
(together with its CP conjugate channel) has been shown to be very promising to simultaneously 
study the $\theta_{13}$ angle and the leptonic CP-violation at the Neutrino Factory 
\cite{Cervera:2000kp,Huber:2002mx} and in the BetaBeam based facilities \cite{Zucchelli:sa}. Anyway most of 
the considerations done apply as well to the time reversal transitions $\nu_\mu\to\nu_e$ (and 
the CP conjugate one) that has been considered in the context of SuperBeam facilities like the 
SPL CERN project \cite{Burguet-Castell:2002qx}. The same strategy will be applied in Sect. 
(\ref{sec:tau}) to the study of the $\nu_e\to\nu_\tau$ transition (the so-called ``silver channel'') 
at a Neutrino Factory. As pointed out in \cite{Donini:2002rm,Autiero:2003fu} this channel 
can become very useful in reducing the number of clone solutions.

\section{Probability vs Number of Events}
\label{some}

Following eq.~(1) of~\cite{Burguet-Castell:2001ez}, the appearance probability for the 
$\nu_e \to \nu_\mu$ ($\bar \nu_e \to \bar \nu_\mu$) transition, expanded at second order 
in perturbation theory in $\theta_{13}$, $\Delta_{12} / \Delta_{23}$, $\Delta_{12}/A$ 
and $\Delta_{12} L$ (see also \cite{Freund:2001pn,Minakata:2002qe}), reads: 
\be 
\label{eq:spagnoli}
P^\pm_{e \mu} (\bar \theta_{13}, \bar \delta) = X_\pm \sin^2 (2 \bar
\theta_{13}) + Y_\pm \cos ( \bar \theta_{13} ) \sin (2 \bar
\theta_{13} ) \cos \left ( \pm \bar \delta - \frac{\Delta_{23} L }{2}
\right ) + Z \, , \ee 
where $\pm$ refers to neutrinos and antineutrinos, respectively. The coefficients $X^\pm, 
Y^\pm$ and $Z$ are defined as follows:
\bea
\label{eq:xcoeff}
X_\pm &=& \sin^2 (\theta_{23} ) 
\left ( \frac{\Delta_{23} }{ B_\mp } \right )^2
\sin^2 \left ( \frac{ B_\mp L}{ 2 } \right ) \ , \\ \label{eq:ycoeff}
Y_\pm &=& \sin ( 2 \theta_{12} ) \sin ( 2 \theta_{23} )
\left ( \frac{\Delta_{12} }{ A } \right )
\left ( \frac{\Delta_{23} }{ B_\mp } \right )
\sin \left ( \frac{A L }{ 2 } \right )
\sin \left ( \frac{ B_\mp L }{ 2 } \right ) \ , \\ \label{eq:zcoeff}
Z &=& \cos^2 (\theta_{23} ) \sin^2 (2 \theta_{12})
\left ( \frac{\Delta_{12} }{ A } \right )^2
\sin^2 \left ( \frac{A L }{ 2 } \right ) \ , 
\eea
with $A=\sqrt{2} G_F n_e$ (expressed in eV$^2$/GeV) and $B_\mp=|A\mp\Delta_{23}|$. Finally, 
$\Delta_{23} = \Delta m^2_{23} / 2E_\nu$ and $\Delta_{12} = \Delta m^2_{12} / 2 E_\nu$. 
Instead of using the coefficients in eqs.~(\ref{eq:xcoeff}-\ref{eq:zcoeff}), it is convenient 
\cite{Migliozzi:2003pw} to introduce the following ($\theta_{13},\delta$)-independent terms:
\be
\label{defo}
O^\pm_1 = X^\pm \, , \quad 
O^\pm_2 = Y^\pm \cos \left ( \frac{\Delta_{23} L}{2} \right ) \, , \quad
O^\pm_3 = Y^\pm \sin \left ( \frac{\Delta_{23} L}{2} \right ) \, , \quad
O_4 = Z \, .
\ee

Notice, however, that transition probabilities, like for example $P^\pm_{e\mu}$, are not the 
experimetally ``measured'' quantities. Experimental results are given in terms of the number 
of charged leptons observed in a specific detector. For the Neutrino Factory Golden channel, 
for example, one measures the number of interacting muons with charge opposite to those 
circulating in the storage ring. In a SuperBeam-driven detector the signal is, on the other 
hand, represented by electrons.
If the considered detector can measure the lepton and hadron final energies, $E_l, E_h$
with a good resolution, the events are then grouped in bins of energy interval $\Delta E_l$. 
This is indeed the case for the proposed Magnetized Iron Calorimeter (MIC) \cite{Cervera:2000vy} 
at the Neutrino Factory, where the number of muons in the i-th energy bin for the input pair 
($\bar \theta_{13},\bar\delta$) and for a parent muon energy $\bar E_\mu$ is given by:
\bea
N^i_{\mu^\mp} (\bar \theta_{13}, \bar \delta)
&=& \left \{ \frac{d \sigma_{\nu_\mu (\bar \nu_\mu)} (E_\mu, E_\nu) }{d E_\mu}
                           \, \otimes \,
                 P^\pm_{e\mu} (E_\nu, \bar \theta_{13}, \bar \delta)           
                           \, \otimes \,
                 \frac{d \Phi_{\nu_e (\bar \nu_e) } (E_\nu, \bar E_\mu)}{d E_\nu} 
                      \right \}_{E_i}^{E_i + \Delta E_\mu} 
\label{eq:convogolden}
\eea
where $\otimes$ stands for a convolution integral \cite{Donini:2002rm}. The same happens for 
the Emulsion Cloud Chamber (ECC) \cite{Autiero:2003fu} (see Sect.~\ref{sec:tau} for details). 
On the contrary, whenever the considered detector energy resolution is not sufficient to 
group events in bins of energy, one single sample will be considered. This is for example 
the case for the Water Cherenkov detector located at the Frejus as the target of a SuperBeam (SB) 
or Beta-Beam (BB) experiments, as we can read from \cite{Zucchelli:sa}.

Using the $O_j$ coefficients in eq.~(\ref{defo}) we define the following integrals: 
\bea
I_j^\pm &=& \left \{ 
    \frac{d \sigma_{\nu_\mu (\bar \nu_\mu)} (E_\mu, E_\nu) }{d E_\mu} \, \otimes \,
            O_j^\pm (E_\nu) \, \otimes \,
    \frac{d \Phi_{\nu_e (\bar \nu_e) } (E_\nu, \bar E_\mu)}{d E_\nu} \right 
            \}_{E_i}^{E_i + \Delta E_\mu} \, ,
\label{eq:convogoldeni}
\eea
where the $O_j$ coefficients are convoluted over the $\nu N$ cross-section and the differential 
neutrino flux. All the experimental details, such as the flux shape, the cross-section, the 
baseline and the solar, atmospheric and matter parameters are encoded in these integrals.
The number of events in the $i$-th bin, in the same approximation used for eqs.~(\ref{eq:spagnoli}), 
is therefore:
\bea
N^i_{\mu^-} &=& \left \{ I^+_1 \sin^2 ( 2 \theta_{13}) + 
                \left [ I^+_2 \cos \delta + I^+_3 \sin \delta \right ]
                \cos \theta_{13} \sin (2 \theta_{13}) +
                I_4 \right \}^i \, , \nn \\
&& \label{eq:numeroeventi} \\
N^i_{\mu^+} &=& \left \{ I^-_1 \sin^2 ( 2 \theta_{13}) + 
                \left [ I^-_2 \cos \delta - I^-_3 \sin \delta \right ]
                \cos \theta_{13} \sin (2 \theta_{13}) +
                I_4 \right \}^i \, . \nn 
\eea
For a specific energy bin and fixed input parameters ($\bar\theta_{13},\bar\delta$), 
we derive the clone location in the ($\theta_{13},\delta$) plane solving:
\bea
\label{eq:ene0}
N^i_{\mu^\pm}(\bar \theta_{13}, \bar \delta; \bar s_{atm}, \bar s_{oct}) &=& 
N^i_{\mu^\pm} ( \theta_{13},  \delta; s_{atm} = \bar s_{atm}, s_{oct} = 
\bar s_{oct})\\ 
\label{eq:ene0t23}
N^i_{\mu^\pm}(\bar \theta_{13}, \bar \delta; \bar s_{atm}, \bar s_{oct}) &=& 
N^i_{\mu^\pm} ( \theta_{13},  \delta; s_{atm} = \bar s_{atm}, s_{oct} = -\bar
s_{oct})\\ 
\label{eq:ene0sign}
N^i_{\mu^\pm}(\bar \theta_{13}, \bar \delta; \bar s_{atm}, \bar s_{oct} )&=&
N^i_{\mu^\pm} ( \theta_{13}, \delta; s_{atm} = -\bar s_{atm}, s_{oct} = \bar
s_{oct})\\  
\label{eq:ene0t23sign}
N^i_{\mu^\pm}(\bar \theta_{13}, \bar \delta; \bar s_{atm}, \bar s_{oct} )&=& 
N^i_{\mu^\pm} ( \theta_{13},  \delta; s_{atm} = -\bar s_{atm}, s_{oct} = -\bar s_{oct})
\eea
as it was the case for the transition probability in eqs.~(\ref{eq:equi0int}-\ref{eq:equi0t23sign}).

\section{Location of the clones}

In this section we determine the analytic expression for the clone location, in the case 
of $\nu_e \to \nu_\mu$ oscillations, solving the integrated systems in 
eqs.(\ref{eq:ene0}-\ref{eq:ene0t23sign}). We are going to describe completely the procedure 
for calculating the {\em intrinsic} clone and then we present, for all the cases, the 
simplified vacuum formul\ae. A detailed derivation is deferred to Appendix \ref{appe}. 
Moreover, in order to simplify the notation, from now on we will omit any reference to 
the bin index $i$ in the convolution integrals, eq.~(\ref{eq:convogoldeni}).

\subsection{The intrinsic clone}
\label{sec:intriclone}

We derive now the theoretical solution of eq.~(\ref{eq:ene0}) following the outline of 
\cite{Donini:2002rm}. This equation corresponds to the particular case $s_{atm} = \bar s_{atm}; 
s_{oct} = \bar s_{oct}$. The explicit results that we get for $\delta$ and $\theta_{13}$ 
can be compared with \cite{Burguet-Castell:2001ez,Donini:2002rm}.

Eq.~(\ref{eq:ene0}) draws two continuous equal-number-of-events (ENE) lines in the ($\theta_{13},\delta$) plane. 
We derive an implicit equation in $\delta$:
\be 
\label{eq:primo}
F^\pm(\delta ) = G^\pm( \theta_{13}, \bar \theta_{13}, \bar \delta ) \, , 
\ee
where 
\be 
\label{eq:secondo}
F^\pm ( \delta ) = \cos \delta \pm \left ( \frac{I^\pm_3}{I^\pm_2} \right ) \sin \delta
\ee
and 
\be
G^\pm ( \theta_{13}, \bar \theta_{13}, \bar \delta) =
\left ( \frac{I^\pm_1}{ I^\pm_2 } \right ) f(\theta_{13},\bar \theta_{13})
+ F^\pm (\bar \delta )\, g( \theta_{13}, \bar \theta_{13} ) \, 
\label{eq:system}.
\ee
The two $\theta_{13}$-dependent functions, $f$ and $g$, are:
\be
\label{eq:intrifunctions}
\left \{
\begin{array}{lll}
f(\theta_{13},\bar \theta_{13}) &=& \dd 
     \frac{\sin^2 (2 \bar \theta_{13}) - \sin^2 (2 \theta_{13})}
          {\cos \theta_{13} \sin (2 \theta_{13})} \\
g(\theta_{13}, \bar \theta_{13}) &=& \dd 
     \frac{\cos\bar\theta_{13}\sin(2\bar\theta_{13})}
          {\cos\theta_{13}\sin(2\theta_{13})} 
\end{array}
\right .
\ee
This parametrization is valid for any value of $\bar \theta_{13}$, including $\bar \theta_{13} \to 0$: 
not an irrelevant feature, since even in the case of vanishing $\bar\theta_{13}$
clone solutions with a non null value of $\theta_{13}$ can appear. In this case
a totally wrong conclusion about the value of $\bar \theta_{13}$ can indeed 
originate by a naive analysis of the experimental results.
Notice, finally, the trivial limits $f(\bar \theta_{13}, \bar \theta_{13}) = 0$ 
and $g (\bar \theta_{13}, \bar \theta_{13}) = 1$.

Solving the system in eq.~(\ref{eq:primo}), we derive two equations for 
$\sin \delta$ and $\cos \delta$:  
\bea
\sin \delta &=& C \, f(\theta_{13}, \bar \theta_{13})+g(\theta_{13}, \bar
\theta_{13})\, \sin \bar \delta \nn \\ & & 
\label{eq:defcd} \\
\cos \delta &=& D \, f(\theta_{13}, \bar \theta_{13})+g(\theta_{13}, \bar
\theta_{13})\, \cos \bar \delta \nonumber
\eea
where the coefficients are
\be
C  =  \dd \left [ \frac{I^+_1 I^-_2 - I^-_1 I^+_2}{
                            I^+_3 I^-_2 + I^-_3 I^+_2} \right ] \, , \qquad
D  = \dd \left [ \frac{I^+_1 I^-_3 + I^-_1 I^+_3}{
                            I^+_3 I^-_2 + I^-_3 I^+_2} \right ]\, .
\label{eq:coeff}
\ee
In the following results, the baseline and energy bin dependence appears only through these 
integral ratios. Notice that only three of the four integrals $I^\pm_j$ are present in $C$ 
and $D$. In fact, there is no dependence on $I_4$ due to the fact that for $s_{oct}=\bar s_{oct}$ 
this integral drops between left and right hand sides of eq.~(\ref{eq:numeroeventi}).

In the approximation $\cos\theta_{13}\simeq \cos\bar\theta_{13} \simeq 1$, perfectly valid 
since $\theta_{13} \leq 10^\circ$ \cite{chooz}, we solve eq.~(\ref{eq:defcd}) for $\theta_{13}$
to get:
\be
\label{eq:solEP}
\sin^2 2 \theta_{13} =
\left \{ 
\begin{array}{l}
\sin^2 2 \bar \theta_{13} \\
\dd \sin^2 2 \bar \theta_{13} + \frac{
1 + 2  \left[ D \cos \bar \delta + C \sin \bar \delta \right] 
\sin 2 \bar \theta_{13} }{ C^2 + D^2 }
\end{array}
\right .
\ee
in which the first solution represents the true input value $\theta_{13}=\bar\theta_{13}$ and 
the second one the location of the intrinsic clone. Eventually, substituting the results of 
eq.~(\ref{eq:solEP}) into $f(\theta_{13},\bar\theta_{13})$ and $g(\theta_{13},\bar\theta_{13})$,
we can get the value of $\delta$ for the intrinsic clone using one of the relations in 
eq.~(\ref{eq:defcd}). 

It is interesting to specialize these results to the case of very short baseline $L$, where matter 
effects can be neglected. In the vacuum approximation, the convolution integrals of the four $O_j$ 
coefficients no longer differ for neutrinos and antineutrinos, $\lim_{A \to 0}\, I^\pm_j = I_j$. 
As a consequence, 
\be
\lim_{A \to 0} C = 0 \, \qquad \qquad  \lim_{A \to 0} D = \frac{I_1}{I_2} 
\ee
and 
\bea
\sin \delta &=& g(\theta_{13}, \bar \theta_{13})\, \sin \bar \delta \nn \\ 
& & \label{eq:defcd0} \\
\cos \delta &=& \frac{I_1}{I_2} \, f(\theta_{13}, \bar \theta_{13})+g(\theta_{13}, \bar
\theta_{13})\, \cos \bar \delta \nonumber
\eea
The location of the intrinsic clone in vacuum is:
\be
\label{eq:thetavacuum}
\sin^2 2 \theta_{13} =
\sin^2 2 \bar \theta_{13} + \left(\frac{I_2}{I_1}\right)^2
\left[ 1 + 2 \, \left(\frac{I_1}{I_2} \right)\cos \bar \delta \,
\sin 2 \bar \theta_{13} \right] \, .
\ee 
Notice that in vacuum the clone location depends on $I_1$ and $I_2$, only: the dependence on 
$I_3$ has also dropped from eq.~(\ref{eq:defcd0}). Eq.~(\ref{eq:thetavacuum}) generalizes the 
results of \cite{Burguet-Castell:2001ez} and \cite{Burguet-Castell:2002qx}, where two regimes 
(``atmospheric'' and ``solar'', for large and small $\bar \theta_{13}$, respectively) are 
discussed independently. In particular, at first order in $I_2/I_1$ our vacuum solution for 
$\theta_{13}$ reproduces eq.~(5) of \cite{Burguet-Castell:2002qx}, valid in the atmospheric 
regime:
\bea
\label{eq:vacuumspagnoli}
\sin^2 2 \theta_{13} & \simeq & \sin^2 2 \bar \theta_{13} + 2 \frac{I_2}{I_1} \cos \bar \delta \, 
                                \sin 2 \bar \theta_{13} \\
\cot \delta & \simeq & - \cot \bar \delta - \frac{I_2}{I_1} \frac{1}{\sin 2 \bar \theta_{13} \, 
                                \sin \bar \delta}
\eea
For values of $\bar \theta_{13}$ large enough, $\delta\sim\pi - \bar\delta + {\cal O} (I_2/I_1)$. 

\subsection{Of other clones}
\label{sec:others}

The general case is represented by eq.~(\ref{eq:ene0t23sign}). In this case, integrals on left 
and right hand side of eq.~(\ref{eq:numeroeventi}) are different and we can no longer factorize 
easily the $\theta_{13}$ dependence. As a consequence, the notation is going to be more involved 
with respect to the intrinsic case. To gain analytical insight of the results, however, we can 
carry on a bit more our mathematical manipulation. 
We label as $\bar I_j$ the true integrals (on left hand side of eq.~(\ref{eq:numeroeventi})) and 
$I_j$ those to be determined in the experiment (on right hand side of eq.~(\ref{eq:numeroeventi})).
We modify the functions in eq.~(\ref{eq:intrifunctions}) as follows: 
\be
\label{eq:genefunctions}
\left \{
\begin{array}{lll}
f_\pm^\prime (\theta_{13},\bar \theta_{13}) &=& \dd 
     \frac{\sin^2 (2 \bar \theta_{13}) - [I^\pm_1/\bar I^\pm_1]\sin^2 (2 \theta_{13})}
          {\cos \theta_{13} \sin (2 \theta_{13})} \, , \\
& & \\
h^\prime (\theta_{13} ) &=& \dd \left ( 1 - I_4/ \bar I_4 \right ) 
                            \frac{1}{\cos \theta_{13} \sin (2 \theta_{13})} \, ,
\end{array}
\right .
\ee
with $f_\pm^\prime (\theta_{13},\bar \theta_{13}) = f (\theta_{13},\bar \theta_{13})$ and 
$h^\prime (\theta_{13}) = 0$ in the intrinsic case.

With these definitions we can rewrite eq.~(\ref{eq:system}) as follows:
\be 
G^\pm ( \theta_{13}, \bar \theta_{13}, \bar \delta) =
  \left [ \frac{\bar I^\pm_1}{I^\pm_2} f_\pm^\prime ( \theta_{13}, \bar \theta_{13}) 
+ \frac{\bar I^\pm_2}{I^\pm_2} \bar F^\pm (\bar \delta) g ( \theta_{13}, \bar \theta_{13}) 
+ \frac{\bar I_4}{I^\pm_2} h^\prime (\theta_{13}) \right ] \, ,
\label{eq:gpm}
\ee
where, clearly,
\be 
\bar F^\pm ( \delta ) = \cos \delta \pm \left ( \frac{\bar I^\pm_3 }{\bar I^\pm_2} \right ) 
\sin \delta \, .
\ee

In Tab.~\ref{tabella} we present the relation between the $\bar I_j$ true integrals and 
those that undergo a discrete $s_{atm}$ and/or $s_{oct}$ transformation.
As it can be seen, the effect of $s_{atm} \to - s_{atm}$ is to flip neutrinos with antineutrinos 
(notice that $I_4$ is unaffected by this change). The change in the $\theta_{23}$ octant
introduces a dependence on $\theta_{23}$ in $I_1$ and $I_4$, leaving $I_{2,3}$ unmodified.
Using Tab.~\ref{tabella}, we can specialize eq.~(\ref{eq:gpm}) for the three ambiguities
and derive analytical results for the {\it sign, octant} and {\it mixed} clones. These results
are typically rather cumbersome and we refer the interested reader to 
Appendix \ref{appe} for the explicit analytical formul\ae~for the different clone locations. 

\begin{table}[h!]
\centering
\begin{tabular}{|c|c|c|c|c|}
\hline
&  {\it Intrinsic:} & {\it Octant:} & {\it Sign:} & {\it Mixed:} \\
& $s_{atm} =   \bar s_{atm}$ & $s_{atm} =   \bar s_{atm}$ & $s_{atm} = - \bar s_{atm}$ & $s_{atm} = - \bar s_{atm}$ \\
& $s_{oct} =   \bar s_{oct}$ & $s_{oct} = - \bar s_{oct}$ & $s_{oct} =   \bar s_{oct}$ & $s_{oct} = - \bar s_{oct}$ \\
\hline
& & & & \\
$I_1^\pm$ &$\bar I_1^\pm$ &  $\cot^2(\theta_{23})\, \bar I_1^\pm$ & $\bar I_1^\mp$ &  $\cot^2(\theta_{23}) \, \bar I_1^\mp$ \\
$I_2^\pm$ &$\bar I_2^\pm$ & $\bar I_2^\pm$  & $ - \bar I_2^\mp$  &  $ - \bar I_2^\mp$ \\
$I_3^\pm$ &$\bar I_3^\pm$ & $\bar I_3^\pm$  & $\bar I_3^\mp$ & $\bar I_3^\mp$\\
$I_4$     &$\bar I_4$ & $\tan^2(\theta_{23})\, \bar I_4$ & $\bar I_4$ & $\tan^2(\theta_{23}) \, \bar I_4$\\
\hline
\end{tabular}
\caption{\it Transformation laws of the integrals $I_j$ under the discrete
ambiguities in the $\theta_{23}$ octant and in the $\Delta m^2_{23}$ sign.} 
\label{tabella}
\end{table}

We specialize here to the vacuum case. 
In this case, since the convolution integrals make no distinction between neutrinos and antineutrinos, 
the transformation properties of Tab.~\ref{tabella} simplify considerably. 
In particular, $I_3$ is unaffected by any of the discrete transformation; $I_2$ changes sign 
when $s_{atm} \to - s_{atm}$ and is unaffected by an octant shift; $I_1$ and $I_4$
are unaffected by changes of $s_{atm}$, whereas get a $\theta_{23}$ dependence for an $s_{oct}$ change.
Notice that for $\theta_{23} = 45^\circ$, only $I_2$ is sensitive to the sign of $\Delta m^2_{23}$.

The vacuum limit of eq.~(\ref{eq:genefunctions}) is given by:
\begin{itemize}
\item {\it Intrinsic and Sign Clone}
\be
\left \{
\begin{array}{lll}
f_\pm^\prime (\theta_{13},\bar \theta_{13}) & \to & f (\theta_{13}, \bar \theta_{13}) \\
& & \label{eq:genefunctionsIS} \\
h^\prime (\theta_{13} ) & \to & 0
\end{array}
\right .
\ee
\item {\it Octant and Mixed Clone}
\be
\left \{
\begin{array}{lll}
f_\pm^\prime (\theta_{13},\bar \theta_{13}) & \to & f_0^\prime (\theta_{13}, \bar \theta_{13}) =  
\dd 
     \frac{\sin^2 (2 \bar \theta_{13}) - \cot^2 \theta_{23} \sin^2 (2 \theta_{13})}
          {\cos \theta_{13} \sin (2 \theta_{13})} \, , \\
& & \label{eq:genefunctionsOM} \\
h^\prime (\theta_{13} ) & \to & h_0^\prime (\theta_{13}) = 
\dd \left ( 1 - \tan^2 \theta_{23} \right ) \frac{1}{\cos \theta_{13} \sin (2 \theta_{13})} \, ,
\end{array}
\right .
\ee
\end{itemize}

Eq.~(\ref{eq:defcd0}) for $\sin \delta$ becomes in vacuum: 
\be
\label{eq:sin0}
\sin \delta = g(\theta_{13}, \bar \theta_{13})\, \sin \bar \delta 
\ee
for all the four clones.
The equation for $\cos \delta$ in vacuum is, on the other hand, different in the four cases: 
\begin{itemize}
\item {\it Intrinsic Clone}: \hspace{0.8cm}
$( \cos \delta )_{\rm int} = \dd \frac{\bar I_1}{\bar I_2} \, f(\theta_{13}, \bar \theta_{13}) + g (\theta_{13}, \bar \theta_{13}) \cos \bar \delta 
$
\item {\it Sign Clone}: \hspace{1.5cm} $(\cos \delta )_{\rm sign} = - ( \cos \delta )_{\rm int}$
\item {\it Octant Clone}: \hspace{1cm}
$(\cos \delta )_{\rm oct} = \dd \frac{\bar I_1}{\bar I_2} \, f_0^\prime (\theta_{13}, \bar \theta_{13}) 
                                 + g (\theta_{13}, \bar \theta_{13}) \cos \bar \delta
                                 + \frac{\bar I_4}{\bar I_2} \, h_0^\prime (\theta_{13}) $
\item {\it Mixed Clone}: \hspace{1.1cm} $(\cos \delta )_{\rm mixed} = - (\cos \delta )_{\rm oct}$
\end{itemize}

Since the only difference 
for intrinsic and sign clones and for octant and mixed ones is a sign in front of the $\cos \delta$ equation, 
we only get two vacuum solutions: 
\begin{itemize}
\item {\it Intrinsic and Sign Clones}
\be
\label{eq:thetavacuum2}
\sin^2 2 \theta_{13} = \left \{ 
\begin{array}{l}
\sin^2 2 \bar \theta_{13} \\
\dd \sin^2 2 \bar \theta_{13} + \left(\frac{\bar I_2}{\bar I_1}\right)^2
\left[ 1 + 2 \, \left(\frac{\bar I_1}{\bar I_2} \right)\cos \bar \delta \,
\sin 2 \bar \theta_{13} \right]
\end{array}
\right .
\ee 
\item {\it Octant and Mixed Clones}
\bea
\sin^2 2 \theta_{13}  &=&  \sin^2 2 \bar \theta_{13} + \label{eq:octavacuum2}\\
& & \nonumber \\
&& \hskip-3.5cm + \left \{ 
       \tan^2 \theta_{23} \, \frac{\bar I_2}{\bar I_1} \, \left [ 
         \cos \bar \delta \sin 2 \bar \theta_{13} + \tan^2 \theta_{23} \, \frac{\bar I_2}{2 \bar I_1} 
                                               \right ] 
- \left ( 1 - \tan^2 \theta_{23} \right ) \left ( \sin^2 2 \bar \theta_{13} - \tan^2 \theta_{23} \, 
                                                                               \frac{\bar I_4}{\bar I_1} \right )
    \right \} \nonumber \\
& & \nonumber \\
&& \hskip-3.5cm \pm \tan^2 \theta_{23} \frac{\bar I_2}{\bar I_1} \, \left \{ 
    \left[ \cos \bar \delta \,\sin 2 \bar \theta_{13} + \tan^2 \theta_{23} \, \frac{\bar I_2}{2 \bar I_1} \right]^2
           - \left ( 1 - \tan^2 \theta_{23} \right ) \left ( \sin^2 2 \bar \theta_{13} - \tan^2 \theta_{23} \, 
                                                                                          \frac{\bar I_4}{\bar I_1} \right )
                                          \right \}^{1/2} \nn
\eea
\end{itemize}
Notice that for the intrinsic and sign clones in vacuum, one solution is the true solution $\theta_{13} = \bar \theta_{13}$
whereas the second solution get shifted with respect to $\bar \theta_{13}$. This is not the case for the octant and mixed
clones that, also in the vacuum limit, will undergo a shift with respect to  $\bar \theta_{13}$ for both solutions.
Eq.~(\ref{eq:thetavacuum2}) can be immediately recovered from eq.~(\ref{eq:octavacuum2}) for $\theta_{23} = 45^\circ$.

The solution for the clone location along the $\delta$ axis can be found using eq.~(\ref{eq:sin0}).
Notice that, due to property of the vacuum transition probabilities under a change in the sign 
of $\Delta m^2_{23}$, the $\delta$ shift for the sign and mixed clones is 
obtained replacing $\delta_{\rm sign} = \pi - \delta_{\rm int}$ and $\delta_{mixed} = \pi - \delta_{oct}$.

\subsection{The $\nu_e \to \nu_\tau$ transition}
\label{sec:tau}

The results of Sect.~\ref{sec:intriclone} and \ref{sec:others} can be straightforwardly extended to the 
$\nu_e \to \nu_\tau$ oscillations. In the same approximation used for eq.~(\ref{eq:spagnoli}), 
we get for the oscillation probability:
\be
\label{eq:etau}
P^\pm_{e \tau} (\bar \theta_{13}, \bar \delta) = 
X^\tau_\pm \sin^2 (2 \bar \theta_{13}) -
Y_\pm \cos ( \bar \theta_{13} ) \sin (2 \bar \theta_{13} )
      \cos \left ( \pm \bar \delta - \frac{\Delta_{23} L }{2} \right ) 
+ Z^\tau \, ,
\ee
with the following coefficients:
\bea
X^\tau_\pm &=& \cos^2 (\theta_{23} ) 
\left ( \frac{\Delta_{23} }{ B_\mp } \right )^2 
\sin^2 \left ( \frac{ B_\mp L}{ 2 } \right ) \ , \\
Z^\tau &=& \sin^2 (\theta_{23} ) \sin^2 (2 \theta_{12}) 
\left ( \frac{\Delta_{12} }{ A } \right )^2 
\sin^2 \left ( \frac{A L }{ 2 } \right ) \, .
\eea
The coefficient $Y_\pm$ is defined in eq.~(\ref{eq:ycoeff}), and it appears
multiplied by a factor $(-1)$ with respect to the same term in eq.~(\ref{eq:spagnoli}). 
The different correlation between $\theta_{13}$ and $\delta$ due to this sign 
has been shown to offer a possible solution of the intrinsic ambiguity at a Neutrino Factory, 
\cite{Donini:2002rm,Autiero:2003fu}.

Eq.~(\ref{eq:convogolden}) must be modified to take into account the $\tau \to \mu$ branching ratio
and the energy distribution of muons proceeding from a $\tau$-decay. 
The number of wrong-sign muons in the detector is therefore: 
\bea
N_{\mu^\mp} (\bar \theta_{13}, \bar \delta)
&=& BR(\tau \to \mu) \, 
   \left \{ \left [ 
      \frac{d N_{\mu^\mp} (E_\mu, E_\tau) }{d E_\mu}                    
                           \, \otimes \,
      \frac{d \sigma_{\nu_\tau (\bar \nu_\tau)} (E_\tau, E_\nu) }{d E_\tau}
           \right ] \right .\nn \\
& & \left .  \qquad             \, \otimes \,
                 P^\pm_{e\tau} (E_\nu, \bar \theta_{13}, \bar \delta)           
                           \, \otimes \,
                 \frac{d \Phi_{\nu_e (\bar \nu_e) } (E_\nu, \bar E_\mu)}{d E_\nu} 
                      \right \}_{E_i}^{E_i + \Delta E_\mu} \nonumber \, .
\eea
The integrals $I^\tau_j$ are consequently: 
\bea
I_j&=& BR(\tau \to \mu) \, 
   \left \{ \left [ 
      \frac{d N_{\mu^\mp} (E_\mu, E_\tau) }{d E_\mu}                    
                           \, \otimes \,
      \frac{d \sigma_{\nu_\tau (\bar \nu_\tau)} (E_\tau, E_\nu) }{d E_\tau}
           \right ] \right .\nn \\
& & \left .  \qquad             \, \otimes \,
                 O_j^\pm (E_\nu)            
                           \, \otimes \,
                 \frac{d \Phi_{\nu_e (\bar \nu_e) } (E_\nu, \bar E_\mu)}{d E_\nu} 
                      \right \}_{E_i}^{E_i + \Delta E_\mu} \, .
\eea
The analytic results of Sect.~\ref{sec:intriclone} and \ref{sec:others} are 
formally extended to the $\nu_e \to \nu_\tau$ oscillation replacing the $I^\mu_j$ with $I^\tau_j$ as 
follows:
\be
\left \{ 
\begin{array}{lll}
I^\mu_1 & \to & \tan^2 \theta_{23} I^\tau_1 \\
I^\mu_2 & \to & - I^\tau_2 \\
I^\mu_3 & \to & - I^\tau_3 \\
I^\mu_4 & \to & \cot^2 \theta_{23} I^\tau_4.
\end{array}
\right .
\ee

%%%%%%%%%%%%%%%%%%%%%%%%%%%%%%%%%%%%%%%%%%%%%%%%%%%%%%%%%%%

\section{The Degeneracies flow}
\label{sec:flow}

In this section we will apply our analytic results to illustrate, from a theoretical 
point of view, where the different clones of the true solution, ($\bar \theta_{13}, 
\bar\delta$), are located in the ($\theta_{13},\delta$) plane for different experiments 
that have been discussed in the literature. 
The goal is to show which combination of experiments is better suited to solve some 
(or all) of the degeneracies. Notice that the existence of unresolved degeneracies 
could, in fact, manifests itself in a complete lost of predictability on the 
aforementioned parameters. It must be noticed that this analysis only provides a useful 
tool to detect the experiment synergies that are most promising to measure these two 
still unknown entries of the PMNS matrix. These results must then be confirmed by a 
detailed analysis in which statistics and systematics of a given experiment combination 
are carefully taken into account. This experimental analysis is well beyond the scope 
of this paper where we only focus in giving a theoretical insight of the problem.

The experiments that we have considered in our analysis are the following: 
\begin{itemize}
\item A Neutrino Factory with 50 GeV muons circulating in the storage ring and two 
      detectors: the first one located at $L = 2810$ Km is designed to look for the golden channel 
      (hereafter labeled as NFG); the second one located at $L = 732$ Km is designed to 
      look for the silver channel (hereafter labeled as NFS). 
      The average neutrino energy is: $<E_{\nu_e,\bar \nu_e}> \simeq 30$ GeV.
\item A SuperBeam with 2 GeV protons and one detector located at $L = 130$ Km to look for 
      $\nu_\mu \to \nu_e$ (hereafter labeled as SB). 
      The average neutrino energy is: $<E_{\nu_\mu}> \simeq 270$ MeV and $<E_{\bar \nu_\mu}> \simeq 250$ MeV.
\item A Beta Beam with ${}^6$He and ${}^{18}$Ne ions and one detector located at $L = 130$ Km to look for 
      $\nu_e \to \nu_\mu$ (hereafter labeled as BB).
      The average neutrino energy is: $<E_{\nu_e}> \simeq 360$ MeV and $<E_{\bar \nu_e}> \simeq 240$ MeV.
\end{itemize}

Notice that for our theoretical analysis the details of the detectors (such as the mass or the 
specific technology adopted to look for a given signal) are irrelevant, since they cancel between 
left and right hand side of eqs.~(\ref{eq:ene0})-(\ref{eq:ene0t23sign}). However, the relevant 
parameters (the neutrino fluxes and the baseline) have been chosen considering the following proposed
experiments: \cite{Cervera:2000kp,Cervera:2000vy} (NFG), \cite{Donini:2002rm,Autiero:2003fu} (NFS), 
\cite{Gomez-Cadenas:2001eu} (SB) and \cite{Zucchelli:sa} (BB). For the different beams we have 
taken as representative the CERN Neutrino Factory, SPL and BetaBeam proposals\footnote{See for 
example in \cite{Apollonio:2002en} for a detailed description of each of these proposals.}. 

The input parameters used in the computation are: 
\begin{itemize}
\item The Atmospheric Sector:
$\Delta m^2_{23} = 2.5 \times 10^{-3} {\rm eV}^2 \, , \, \theta_{23} = 40^\circ $;
\item The Solar Sector: 
$\Delta m^2_{12} = 7.3 \times 10^{-5} {\rm eV}^2 \, , \, \theta_{12} = 35^\circ $;
\item The Matter Parameter: $A = 1.1 \times 10^{-4} \, {\rm eV}^2 / {\rm GeV}$ .
\end{itemize}

In Sect.~\ref{sec:eqene} we compare the clone locations derived from equiprobability and ENE curves;
in Sect.~\ref{sec:flow1} we consider the intrinsic and the sign clones; in Sect.~\ref{sec:flow2} 
we focus on the octant and the mixed clones. We have shown in Sect.~\ref{sec:others} that the two 
pairs of solutions are indeed related in the vacuum limit.

\subsection{Equiprobability flow vs ENE flow}
\label{sec:eqene}

From a theoretical point of view one can look at the clone displacement with respect to the 
true solution ($\theta_{13} = \bar\theta_{13}$, $\delta = \bar\delta$) solving either the 
systems for the equiprobability curves, eqs. (\ref{eq:equi0int}-\ref{eq:equi0t23sign}), 
or the corresponding systems for the ENE curves, eqs.~(\ref{eq:ene0}-\ref{eq:ene0t23sign}). 
One should expect the two treatments to be equivalent: this is not always the case.  

\begin{figure}[t]
\begin{center}
\begin{tabular}{cc}
\hspace{-0.55cm} \epsfxsize8cm\epsffile{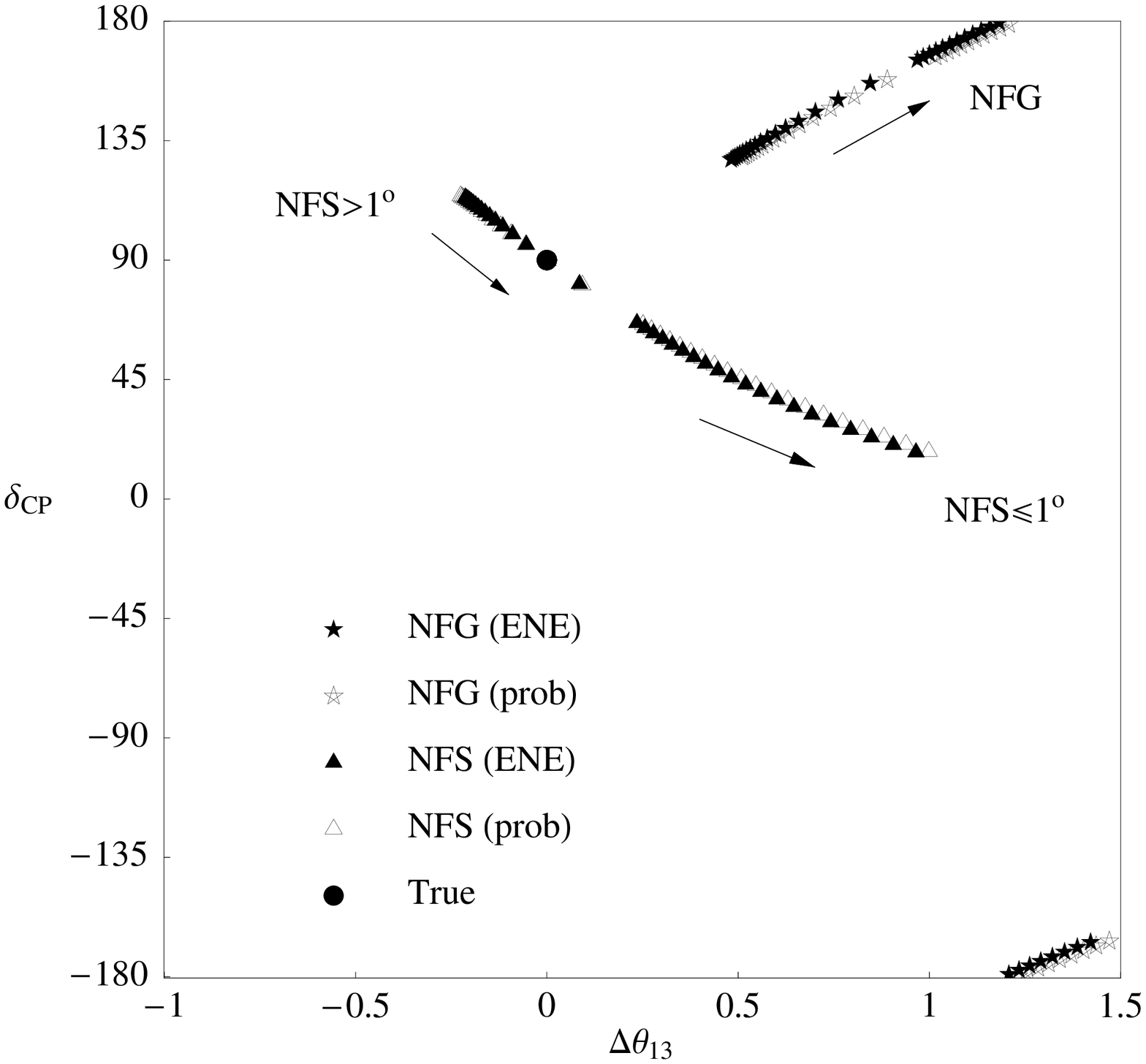} 
                \hspace{-0.3cm} & \hspace{-0.3cm}
                \epsfxsize8cm\epsffile{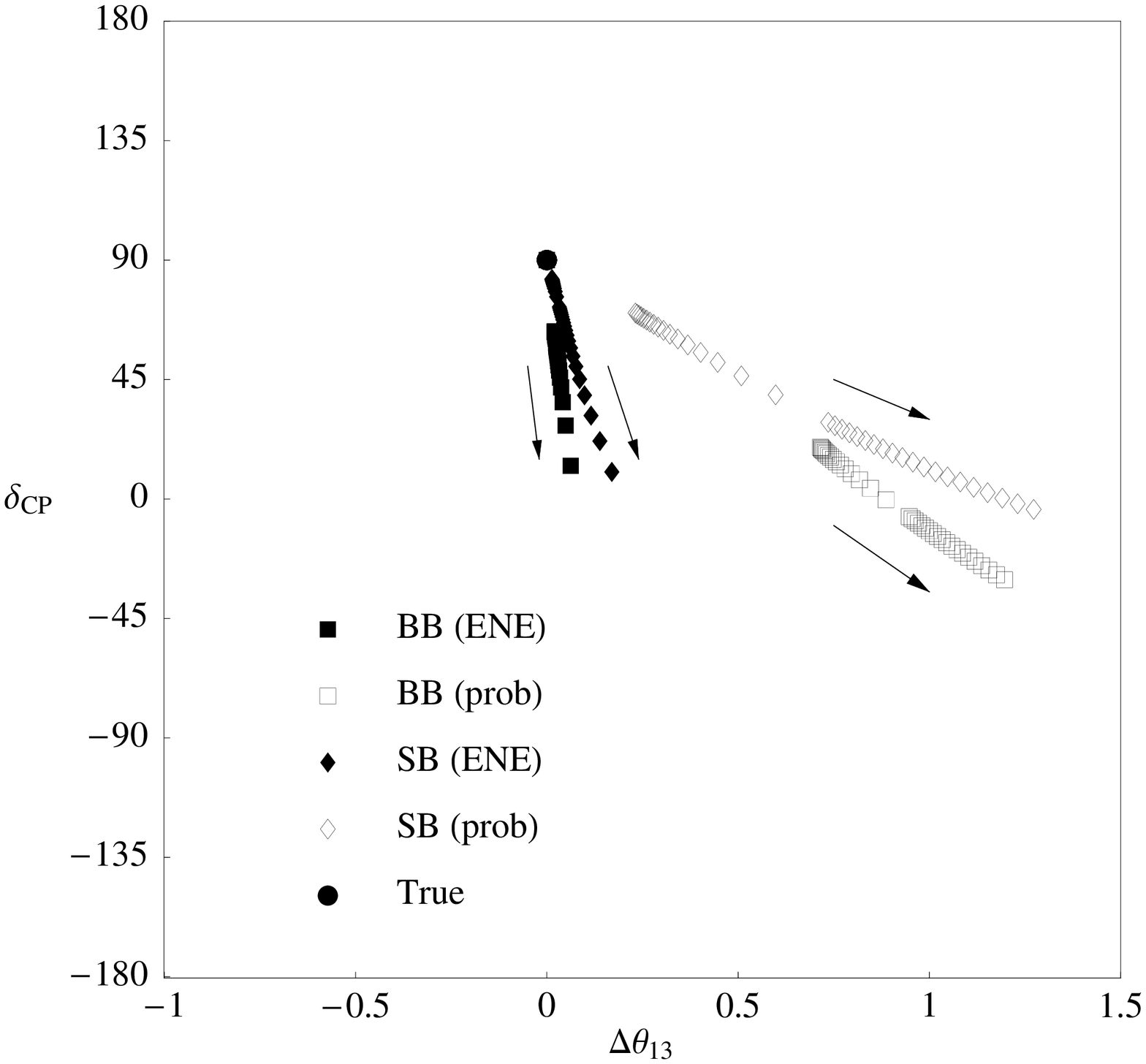}
\vspace{-0.4cm}
\end{tabular}
\caption{\it 
Intrinsic clone flow for the Neutrino Factory golden and silver channels (left)
and the SuperBeam and BetaBeam signals (right), in the ($\Delta \theta_{13},\delta$) 
plane for $\bar \theta_{13} \in [0.1^\circ,10^\circ], \bar \delta = 90^\circ$. 
The full circle is the true solution. The full (empty) symbols represent
the intrinsic clone flow, computed from the ENE (equiprobability) curves.}
\label{fig:intriflow}
\end{center}
\end{figure}

In Fig.~\ref{fig:intriflow} we plot the location of the intrinsic clone that results 
from the equiprobability curves and from the ENE curves. These points graphically depict 
how far from the true solution the degeneracy falls. A small change in the input parameter 
$\bar\theta_{13}$ results in a small shift of the clone location (remind, however, the 
$2\pi$-periodicity in the $\delta$ axis). Almost continuous geometrical regions where 
degeneracies lie are therefore defined for a given interval in $\bar \theta_{13}$, 
illustrating how the clones move due to a change in the input parameters: we will call this 
the ``clone flow''. In the left plot we present the results for the Neutrino Factory-based 
golden and silver channels (i.e. $\nu_e \to \nu_\mu, \nu_\tau$), whereas in the right plot 
those of a SuperBeam-based $\nu_\mu \to \nu_e$ and of a BetaBeam-based $\nu_e \to \nu_\mu$ 
signals. The clone locations have been computed for variable input parameter $\bar\theta_{13}$ 
in the range $\bar \theta_{13} \in [0.1^\circ,10^\circ]$ and fixed $\bar\delta=90^\circ$. 
The arrows describe the direction of the flows: from large to small $\bar\theta_{13}$.
The plots should be read as follows: 
\begin{itemize}
\item 
the horizontal axis is $\Delta\theta_{13}=\theta_{13}-\bar\theta_{13}$. This is the value 
of the $\theta_{13}$-shift, as derived in the second solution of eq.~(\ref{eq:solEP});
\item
the vertical axis is the CP-violating phase $\delta$ itself, as it can be derived
introducing the result of eq.~(\ref{eq:solEP}) into one of eqs.~(\ref{eq:defcd}).
\end{itemize}

For any value of the input parameter $\bar\theta_{13}$ the true solution is represented 
by the full (black) circle in $\Delta \theta_{13} = 0$, $\delta=\bar \delta = 90^\circ$. 
Notice that this property is specific to the intrinsic case. In general, both solutions of
eqs.~(\ref{eq:ene0t23})-(\ref{eq:ene0t23sign}) will differ from the true solution 
($\bar \theta_{13}, \bar \delta$). The intrinsic clones, computed using the ENE or the 
equiprobability equations, are respectively plotted as full or empty symbols as explained 
in the legend and in the caption. 

In the left plot of Fig.~\ref{fig:intriflow} we can clearly identify the golden and silver channel
intrinsic clone flow\footnote{For this specific plot 
we are choosing the energy bin between 30-40 GeV but the same conclusion holds for all 
the other bins as well for the integrated flux.}. 
The NFG flow goes from the true solution toward positive $\Delta \theta_{13}$ and 
positive $\delta - \bar \delta$: for $\bar \theta_{13} = 10^\circ$ we have $\Delta \theta_{13} \sim 0.5^\circ$
and for very small $\bar \theta_{13}$, $\Delta \theta_{13} \to 1.5^\circ$. The NFS flow goes from negative $\Delta \theta_{13}$
and positive $\delta - \bar \delta$ toward positive $\Delta \theta_{13}$ and negative $\delta - \bar \delta$. 
The different behavior of the two flows reflects the results of eq.~(\ref{eq:solEP}). For $\nu_e \to \nu_\mu$
transitions, the second solution of eq.~(\ref{eq:solEP}) gives $\sin^2 2 \theta_{13} \ge \sin^2 2 \bar \theta_{13}$
for any value of $\bar \theta_{13}$, being $C$ and $D$ strictly positive quantities. On the contrary, 
due to the formal replacement of the convolution integrals $I_2^\mu \to - I^\tau_2$,$I^\mu_3 \to -I^\tau_3$
we find that $C^\tau,D^\tau$ are strictly negative quantities. For changing $\bar \theta_{13}$, therefore, 
the $\theta_{13}$-shift can either be positive or negative.
Notice how the intrinsic flows computed from the integrated ENE curves and from the equiprobability curves 
(almost) coincide for the Neutrino Factory-based samples.

In the right plot of Fig.~\ref{fig:intriflow} SuperBeam and BetaBeam-based flows are shown. 
The ENE clone flow for the SB facility (full diamonds) has a very small $\Delta\theta_{13}$ ($\le 0.1^\circ$) 
and it is practically insensitive to the change of $\bar\theta_{13}$. 
The shift in $\delta-\bar\delta$ is always negative and become smaller for larger $\bar\theta_{13}$. 
The ENE clone flow for the BB facility (full squares) is practically identical to the SB one. 
This means that the presence of the intrinsic clone in these two facilities does not interfere 
with the measure of $\theta_{13}$, whereas can drastically reduce the ability of measuring the CP phase. 

The important feature is that for all these facilities the difference between using ENE (full symbols) 
or equiprobability (empty symbols) curves is quite substantial. Using theoretical information from the equiprobability 
curves could lead to largely wrong conjectures, in particular regarding the $\bar\theta_{13}$ 
dependence that is practically absent in the ENE case and is conversely quite large 
using the equiprobability curves. The reason why SB and BB present this large difference 
between the two treatments has to be searched in their quite narrow energy spectrum. 
For very peaked spectra the approximation of ``constant flux'', that is 
implicit in the equiprobability curves, badly fails. On the other hand, the broad neutrino spectrum 
from a Neutrino Factory makes the two treatments practically equivalent. 

The theoretical results obtained for the intrinsic clone are quite general: 
a large difference in computing the clone location from ENE or equiprobability curves
is observed for narrow band beam experiments, whereas no difference
is observed for broad band beam experiments. For this reason, in the following sections
we will compute the clone location for the different experiments using the integrated ENE curves.

\subsection{The Intrinsic and the Sign Clone Flow}
\label{sec:flow1}

We now look for the different combinations of proposed facilities that help in solving the eightfold-degeneracy. 
The theoretical requirement for solving ambiguities is to find a specific combination of experiments
such that the corresponding clone flows lie ``well apart''. In this scenario an experimental fit to the data
will result in a good $\chi^2$ absolute minimum near the true solution, where the information from different experiments 
adds coherently. Near the theoretical clone locations poor $\chi^2$ relative minima will be found,
since in these points the results of different experiments do not add coherently. 
\begin{figure}[t!]
\vspace{-0.5cm}
\begin{center}
\begin{tabular}{cc} 
\hspace{-0.55cm} \epsfxsize8cm\epsffile{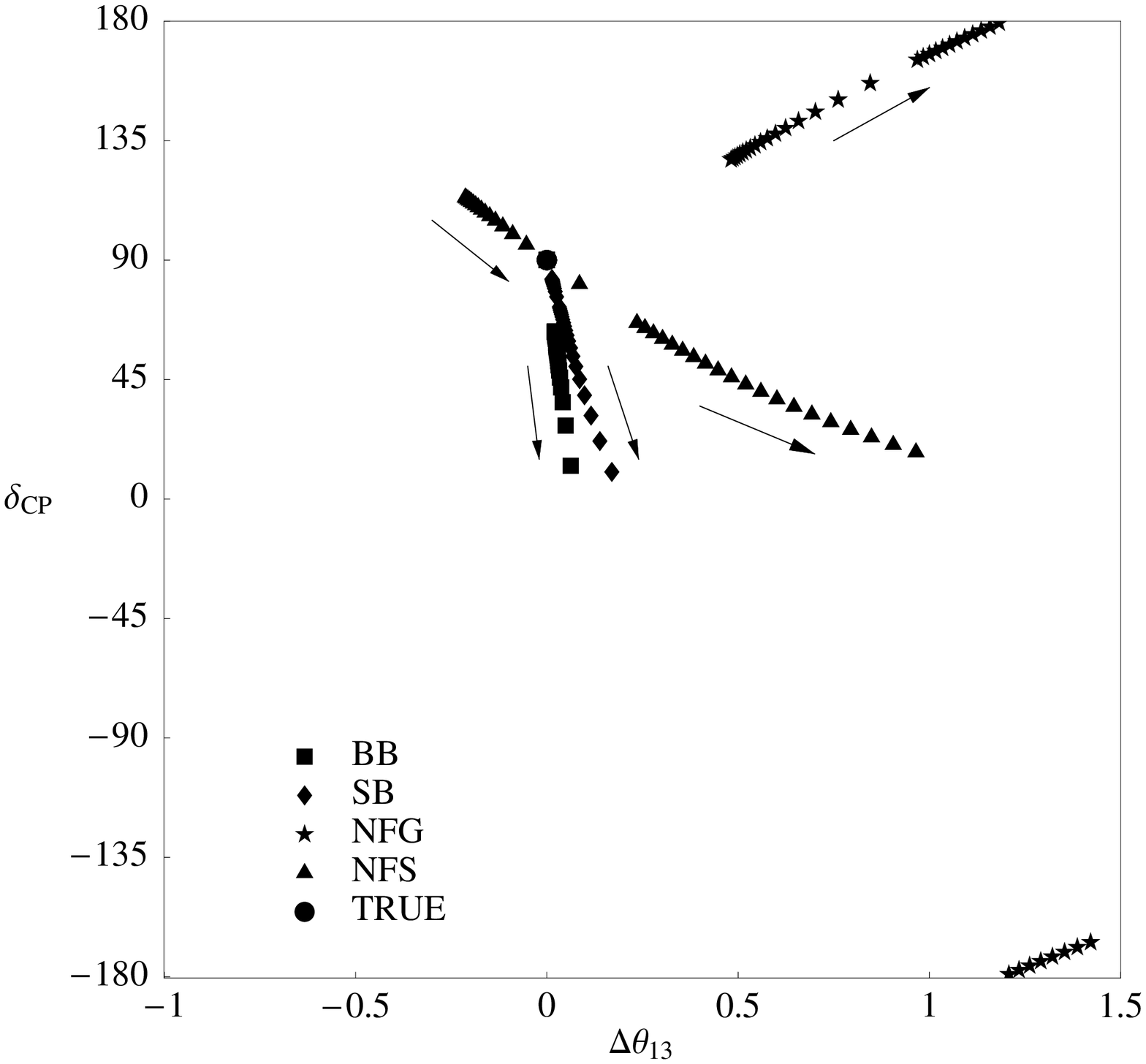} \hspace{-0.3cm} & \hspace{-0.3cm} 
              \epsfxsize8cm\epsffile{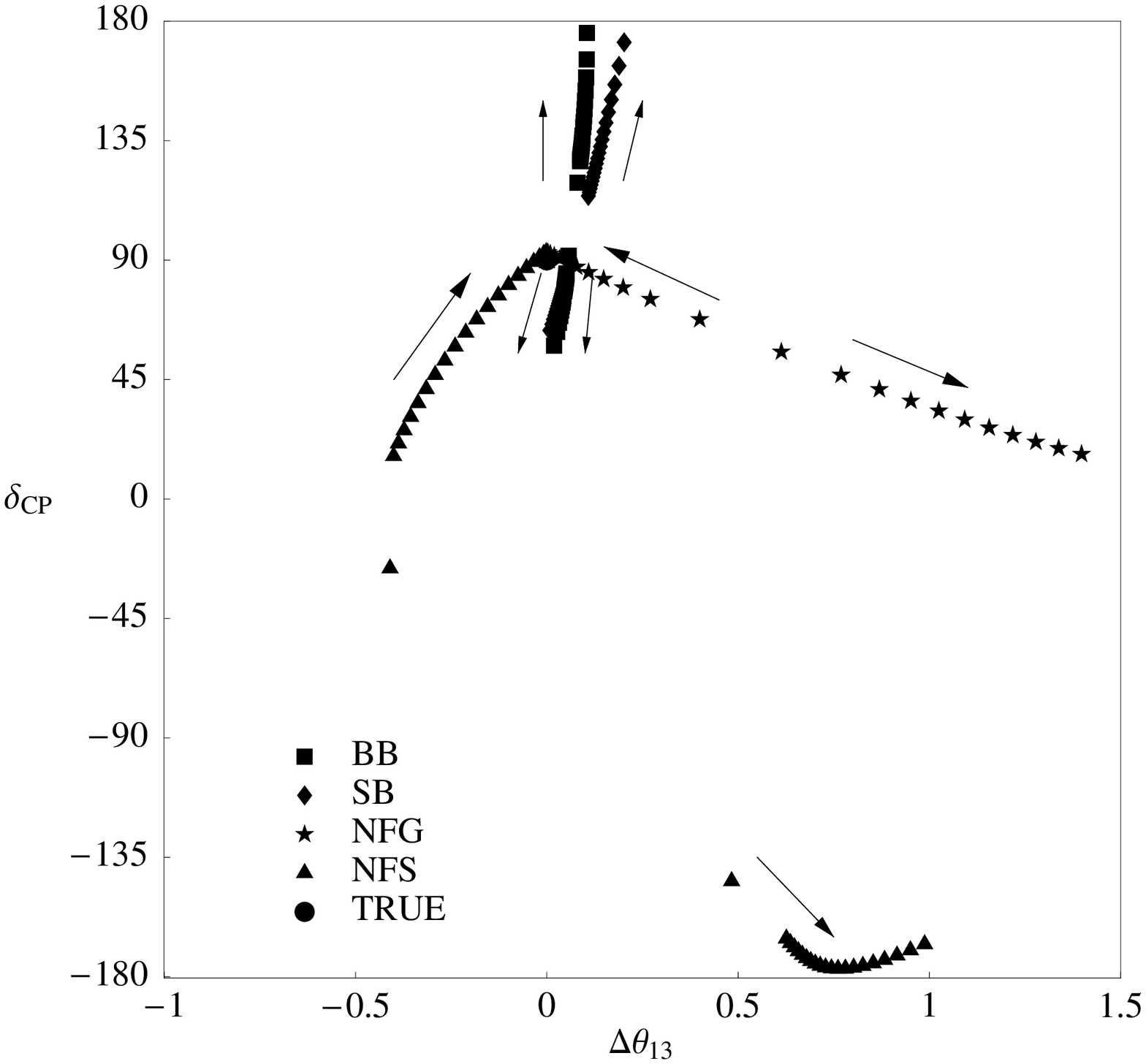} \vspace{-0.5cm}\\ 
\hspace{-0.55cm} \epsfxsize8cm\epsffile{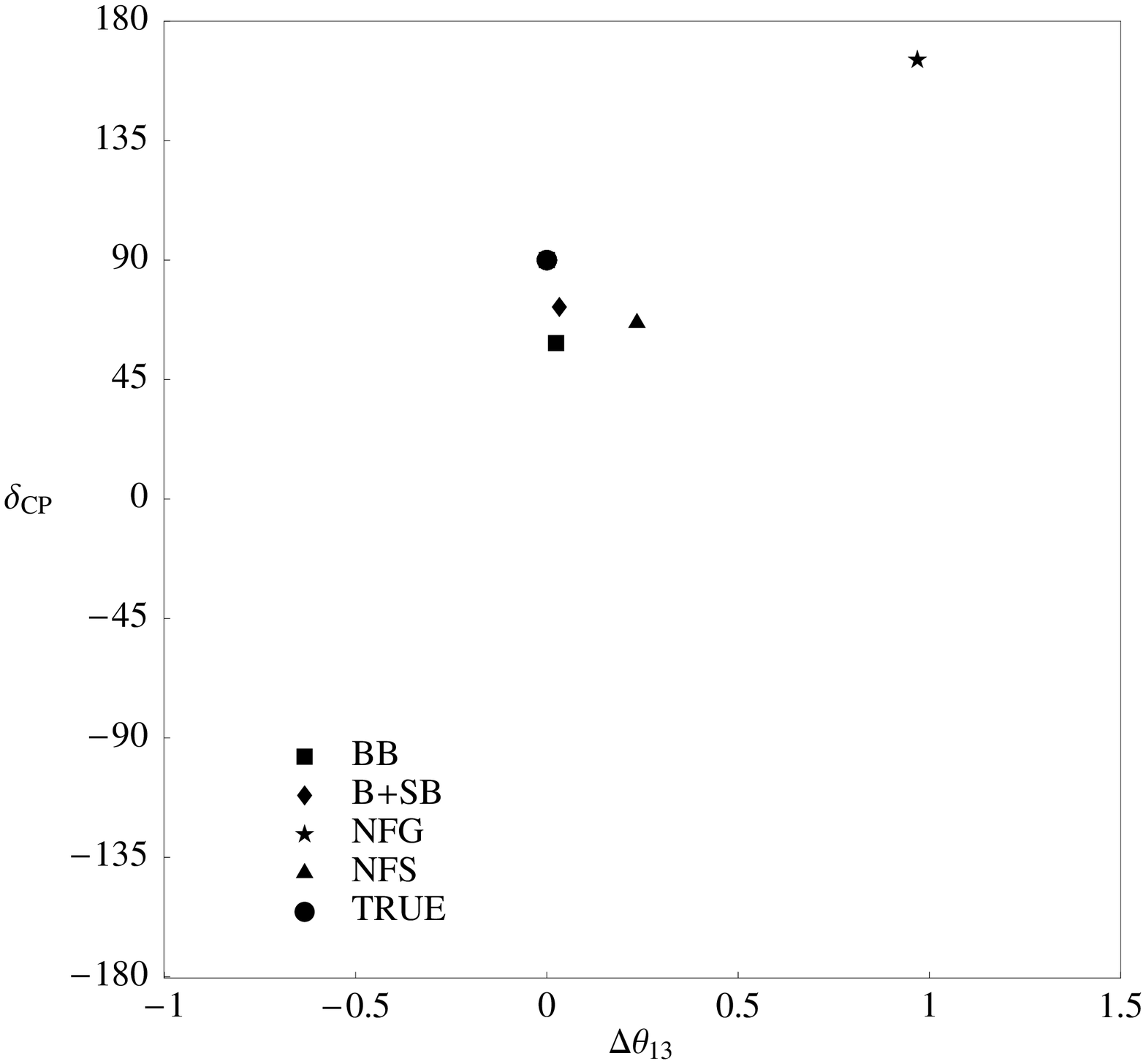} \hspace{-0.3cm} & \hspace{-0.3cm} 
              \epsfxsize8cm\epsffile{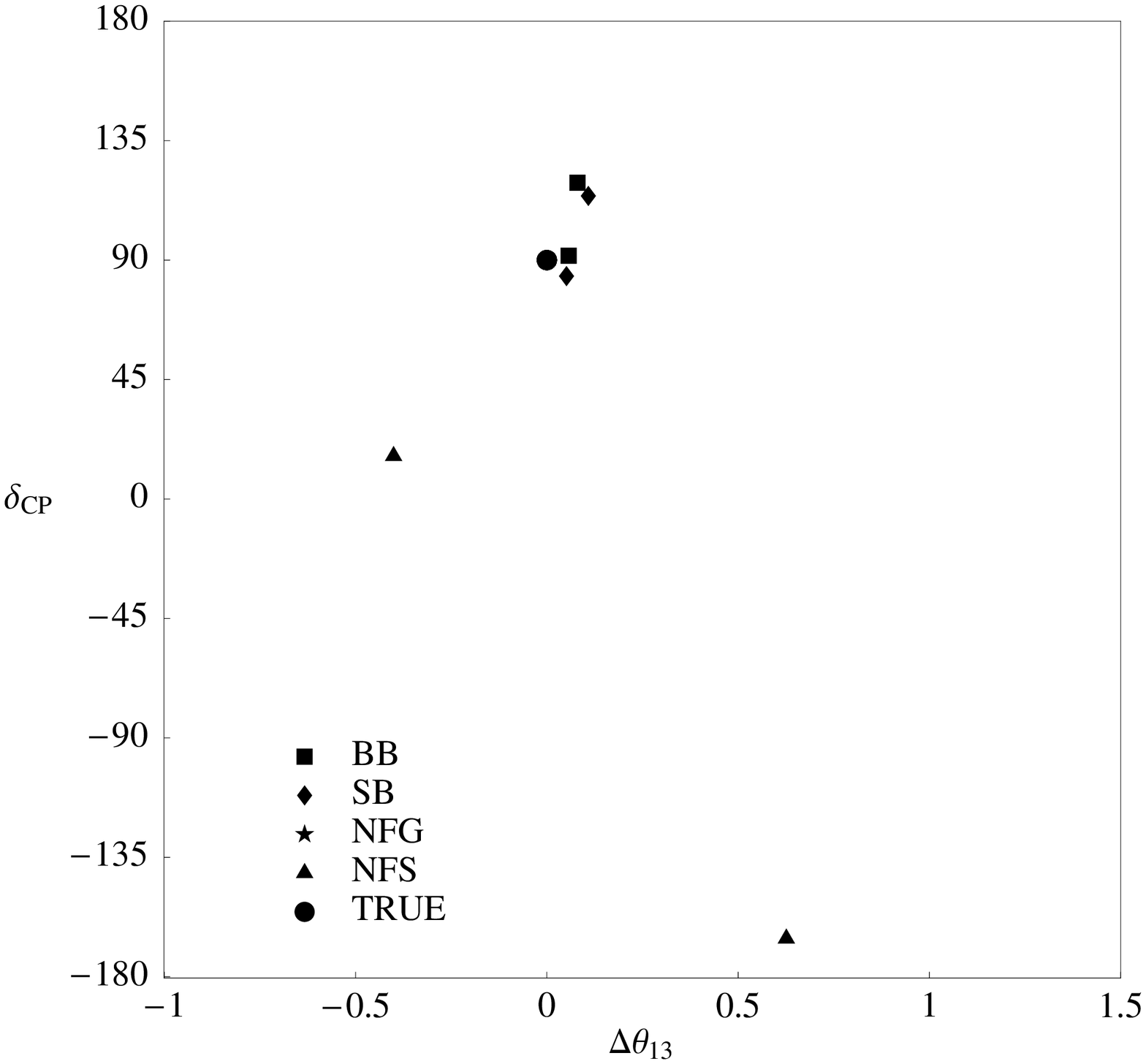} \\
\end{tabular}
\caption{\it 
Clone location in the ($\Delta\theta_{13},\delta$) plane for the intrinsic and sign degeneracies: 
BB (boxes); SB (diamonds); NFG (stars) and NFS (triangles).
The full circle is the true solution, $\Delta \theta_{13} = 0, \bar \delta = 90^\circ$. 
Pictures (a) and (b) represent the intrinsic and sign clone flows 
for $\bar\theta_{13} \in [0.1^\circ,10^\circ]$ and fixed $\bar \delta$. 
Pictures (c) and (d) represent the intrinsic and sign clone location for 
$\bar \theta_{13} = 1^\circ$.}
\label{fig:allflows1}
\end{center}
\end{figure}
In Fig.~\ref{fig:allflows1} we present the clone flow for the intrinsic (left) and the sign (right)
degeneracies. The clone locations have been computed for variable input parameter $\bar\theta_{13}$ 
in the range $\bar \theta_{13} \in [0.1^\circ,10^\circ]$ and fixed $\bar\delta=90^\circ$ (upper 
plots) or for a fixed value, $\bar \theta_{13} = 1^\circ, \bar \delta = 90^\circ$ (lower plots). 
Again the arrows indicate the direction of the flows from large to small $\bar\theta_{13}$.
In each figure the results for the four facilities are presented together, in order to show how 
the combination of two or more of them can help solving that specific degeneracy. 
We illustrate the outcome of our analysis by studying the two degenerations one by one. 

\begin{itemize}
\item {\it The Intrinsic Clone}: Fig.~\ref{fig:allflows1}a/\ref{fig:allflows1}c. 

The SB and BB intrinsic clone flows are extremely similar. This has already been noticed in 
the previous section, and is the result of the beam design. In \cite{Zucchelli:sa} has been 
proposed to run the SPL and the BetaBeam at the same time and using the same detector located 
at $L=130$ Km. In view of solving the degeneracies, this is clearly not a good option since 
the two flows always lie very near. The combination of these two facilities will still be 
affected by the intrinsic ambiguity. 

This is not the case for the Neutrino Factory setup: the NFG channel is well separated from the 
NFS channel for any value of $\bar\theta_{13}$. The golden channel combined with the silver and/or 
the SB/BB facilities is always, in principle, capable to solve the intrinsic ambiguity (i.e. 
provided that the corresponding experimental samples are all statistically significant and that 
systematic errors under control).

In Fig.~\ref{fig:allflows1}c we plot the intrinsic clone location for 
the small value $\bar \theta_{13} = 1^\circ$, on the verge of the sensitivity limit.

\item {\it The Sign Clone}: Fig.~\ref{fig:allflows1}b/\ref{fig:allflows1}d. 

 It can be seen that again the SB and BB flows lie very near. A combination of them is therefore
not suited to solve the sign degeneracy. On the other hand the Golden and Silver Neutrino Factory flows are 
quite separated. There could be some superposition between all the four flows for certain values of 
$\bar\theta_{13}$ (e.g. NFG and NFS flows in the limit of vanishing $\bar \theta_{13}$). 
Nevertheless this superposition occurs in a region nearby the true solution where the existence 
of a clone is somewhat irrelevant. 
The combination of either NFG or NFS with the SB or the BB signals should easily solve the ambiguity\footnote{
Notice, however, that for large $\bar\theta_{13}$ angles even the Golden channel 
alone can solve the sign ambiguity being strongly sensitive to matter effects.}.
In Fig.~\ref{fig:allflows1}c we plot the sign clone location for $\bar \theta_{13} = 1^\circ$.
Notice that for this specific case no NFG clones were found. This can be interpreted as 
a consequence of the very good (theoretical) sensitivity to matter effects (i.e. the $\Delta m^2_{23}$ sign) of 
the NFG for $L = 2810$ Km. Be aware, however, that in a realistic experimental case a sign clone could appear
also for the NFG due to statistical and systematical uncertainties.
\end{itemize}

As a final comment notice that the results for the sign clone for the SB and BB flows are 
in perfect agreement with the theoretical analysis of Sect. \ref{sec:others}. For these two 
facilities the vacuum limit can be considered a good approximation (being $L = 130$ Km) and 
we therefore expect the same $\Delta \theta_{13}$ shift as in the intrinsic case with 
$\delta_{\rm sign} = \pi - \delta_{\rm int}$. This is indeed the caseby looking at Fig. 
\ref{fig:allflows1}a and Fig. \ref{fig:allflows1}b.

\subsection{The Octant and the Mixed Clones}
\label{sec:flow2}

\begin{figure}[t!]
\vspace{-0.5cm}
\begin{center}
\begin{tabular}{cc} 
\hspace{-0.55cm} \epsfxsize8cm\epsffile{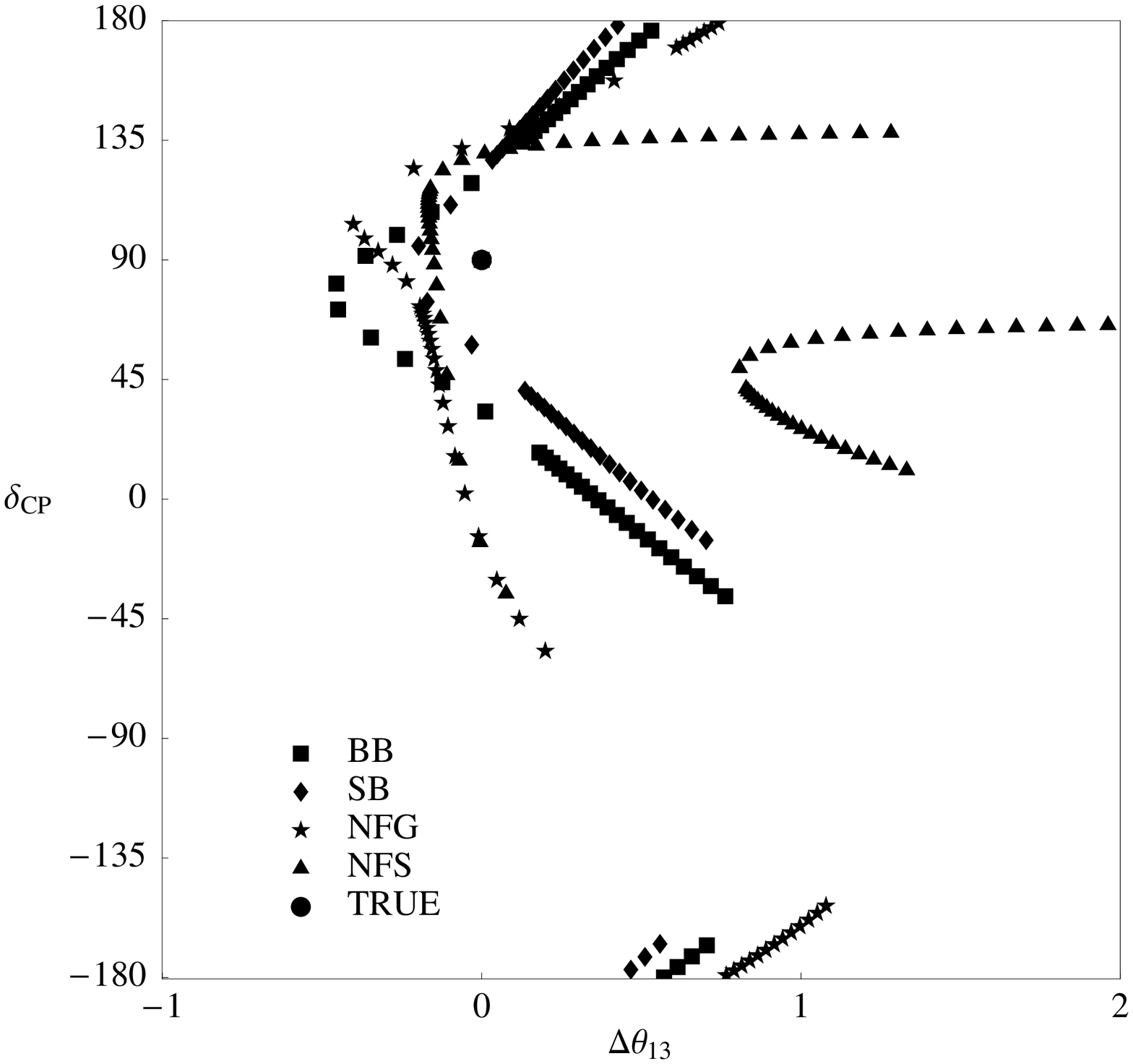} \hspace{-0.3cm} & \hspace{-0.3cm} 
              \epsfxsize8cm\epsffile{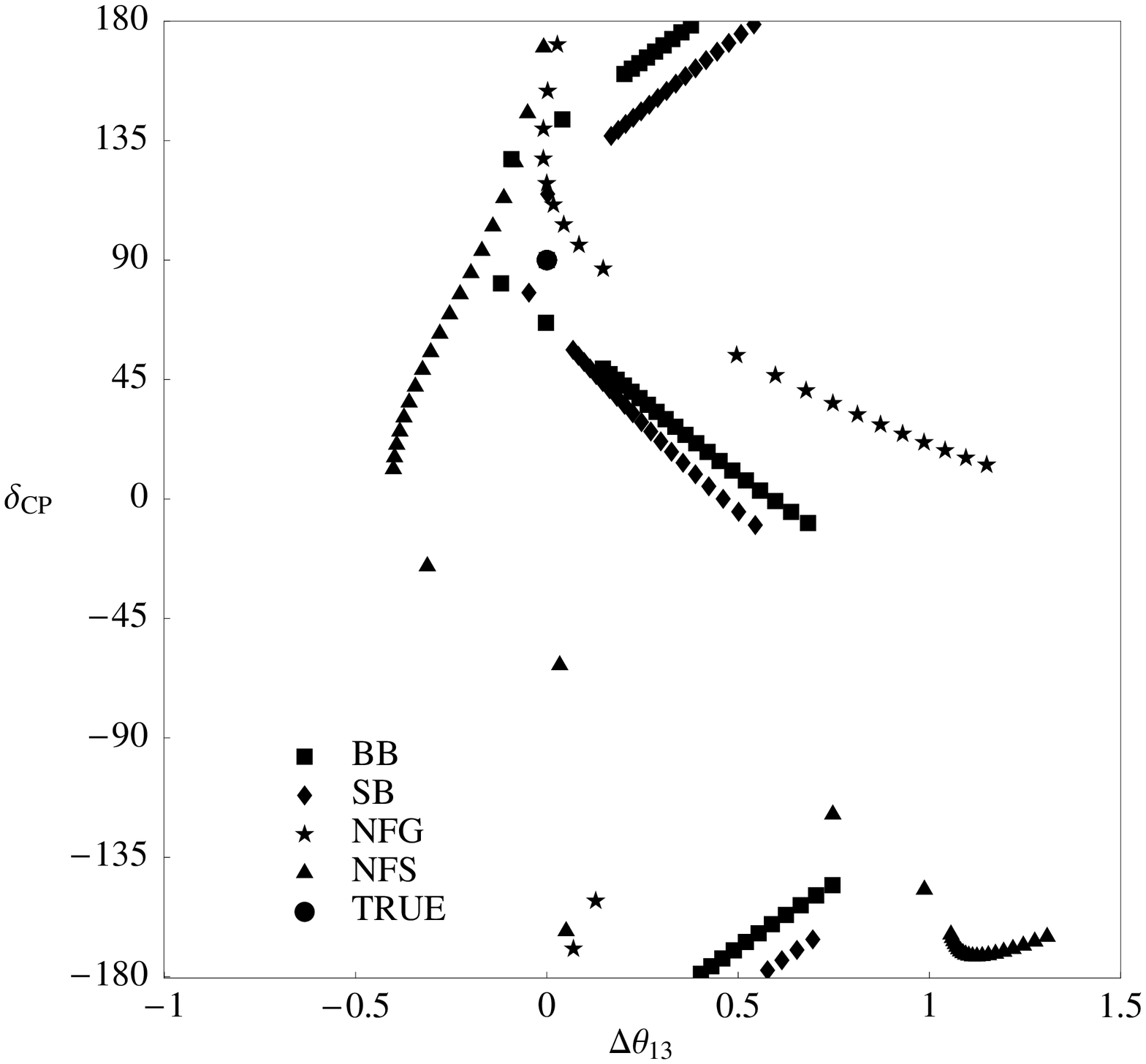} \vspace{-0.5cm} \\
\hspace{-0.55cm} \epsfxsize8cm\epsffile{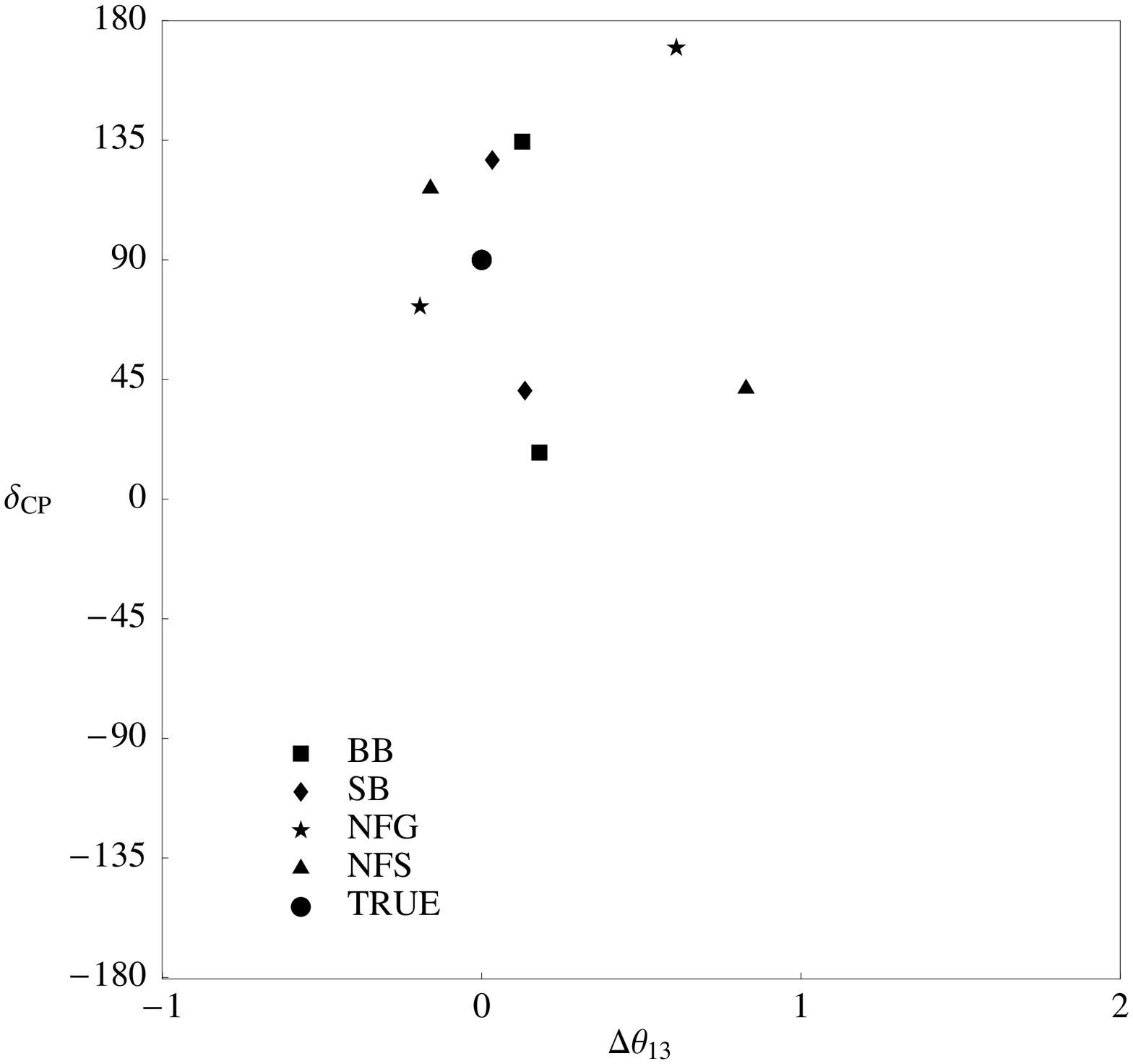} \hspace{-0.3cm} & \hspace{-0.3cm} 
              \epsfxsize8cm\epsffile{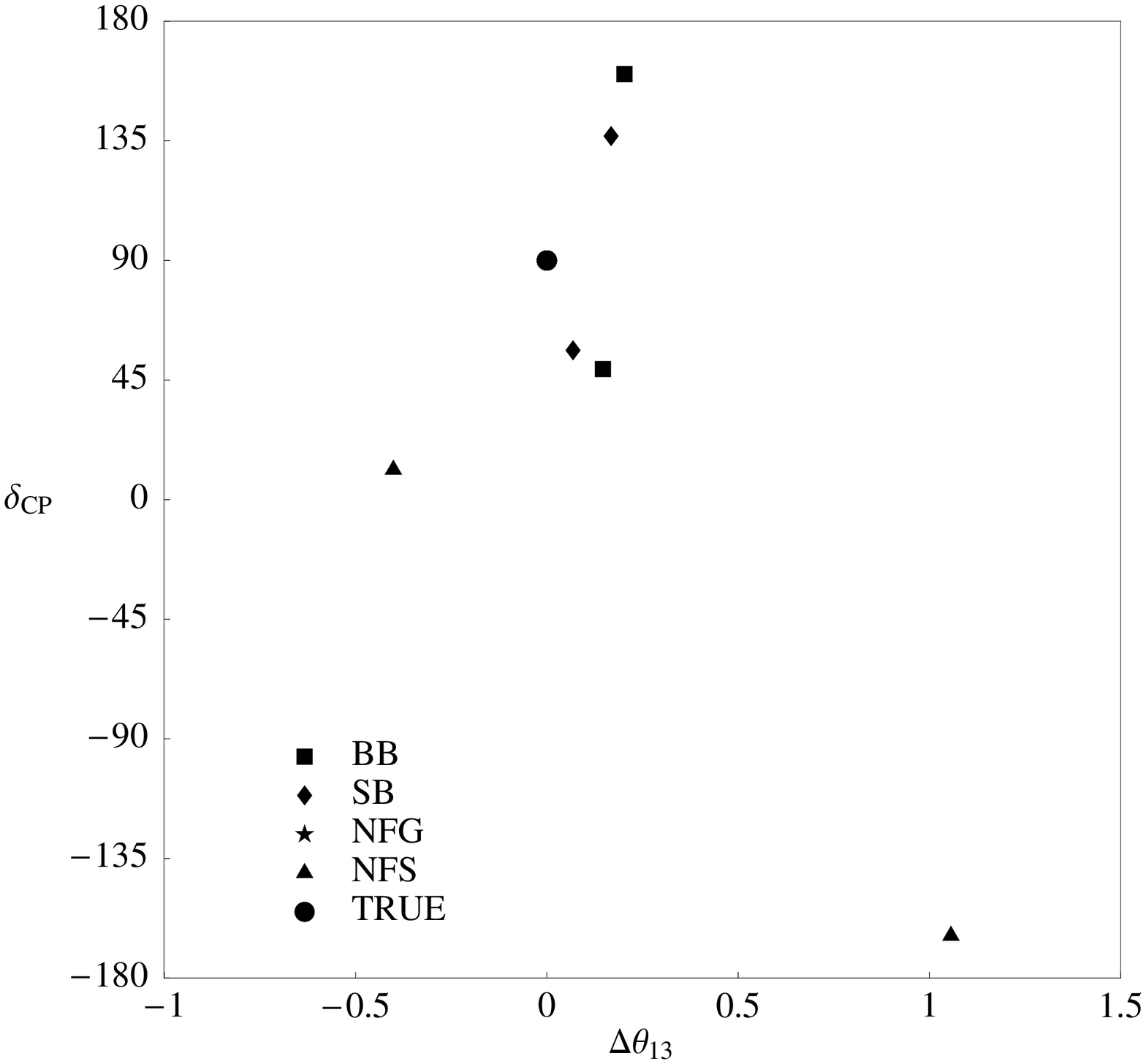} \\
\end{tabular}
\caption{\it 
Clone location in the ($\Delta\theta_{13},\delta$) plane for the octant and mixed degeneracies: 
BB (boxes); SB (diamonds); NFG (stars) and NFS (triangles). The full circle is the true solution, 
$\Delta \theta_{13} = 0, \bar \delta = 90^\circ$. Pictures (a) and (b) represent the octant and 
mixed clone flows for $\bar\theta_{13} \in [0.1^\circ,10^\circ]$ and fixed $\bar \delta$. 
Pictures (c) and (d) represent the octant and mixed clone location for $\bar\theta_{13}=1^\circ$. 
The results have been computed with $\theta_{23}=40^\circ$.}
\label{fig:allflows2}
\end{center}
\end{figure}

If $\theta_{23}$ was maximal no additional degeneracies were present. Otherwise two 
furtherdegeneracies do appear\footnote{Be aware that these clones will appear also for 
maximal $\theta_{23}$ once including the unavoidable experimental error band.}. In 
Fig.~\ref{fig:allflows2} we present the clone flow for the octant (left) and the mixed 
(right) degeneracies. The clone locations have been computed for variable input parameter 
$\bar\theta_{13}$ in the range $\bar \theta_{13} \in [0.1^\circ,10^\circ]$ and fixed 
$\bar\delta=90^\circ$ (upper plot) or for a fixed value, $\bar \theta_{13} = 1^\circ, 
\bar \delta = 90^\circ$ (lower plots). In each figure the results for the four facilities 
are presented together.

In Figs.~\ref{fig:allflows2}a and \ref{fig:allflows2}b a very complicated pattern of clone 
flows appears. For this reason we discuss the specific case of $\bar \theta_{13} = 1^\circ$ 
depicted in Figs.~\ref{fig:allflows2}c and \ref{fig:allflows2}d.

\begin{itemize}
\item {\it The Mixed Clone}: Fig.~\ref{fig:allflows2}b/\ref{fig:allflows2}d. 

We again notice that BB (boxes) and SB (diamonds) clones lie very near. 
As in the case of the sign ambiguity, no NFG mixed clones appear for this specific $\bar \theta_{13}$.
Finally, the NFS clones are well separated from the BB/SB pairs. 
The combination of the golden and/or the silver channel and one of the BB/SB facilities can, in principle, 
solve the mixed degeneracy.

\item {\it The Octant Clone}: Fig.~\ref{fig:allflows2}a/\ref{fig:allflows2}c. 

This is certainly the most difficult degeneracy to solve, since the flows of the 
four facilities overlap in a significant region of the ($\Delta \theta_{13}, \delta$) plane, 
Fig.~\ref{fig:allflows2}a. It seems feasible to reduce the overall ambiguity by combining NFG 
and NFS or BB/SB facilities, as it can be seen in Fig.~\ref{fig:allflows2}c.
A complete elimination of this degeneracy can perhaps be achieved combining the NFG, NFS
and one of the BB/SB experiments.
\end{itemize}

%%%%%%%%%%%%%%%%%%%%%%%%%%%%%%%%%%%%%%%%%%%%%%%%%%%%%%%%%%%

\section{Conclusions}
\label{sec:concl}

In this paper we derive the theoretical location of the so-called ``clones'' that
appear in the simultaneous measurement of $\theta_{13}$ and $\delta$. The general 
technique used to write analytical formul\ae~ for the clone location for the 
four degeneracies (called {\it intrinsic, sign, octant} and {\it mixed})
has been reported. Our analytical results apply to equiprobability
and equal-number-of-events (ENE) curves as well. However, we have shown that 
the clone location computed from equiprobability or ENE curves can differ significantly
for narrow band beam experiments. Notice that the latter approach reproduces the experimental 
situation, where the signal is represented by charged leptons in the detector 
and not by the oscillation probability itself.
Using theoretical information from the equiprobability curves could thus lead 
to largely wrong conjectures. In the actual calculation we have therefore presented 
results obtained using the ENE curves.

We present the clone flow for experiments fuelled by three different beams,
using as a reference the CERN proposal for a SuperBeam, a BetaBeam and 
a Neutrino Factory. In the case of the Neutrino Factory two different detectors
have been considered to take care of the ``golden'' and ``silver'' channel.

Our results show that the combination of these specific SuperBeam 
and BetaBeam does not help in solving any of the degeneracies.
All the clone solutions for these two facilities lie very near, as a consequence of the fact that 
they have a very similar energy spectrum and that both are using the same detector
located at $L = 130$ Km.

The combination of the Neutrino Factory golden and silver channel, on the contrary, 
seems to help significantly to reduce ambiguities. In our theoretical analysis, indeed, 
the intrinsic, sign and mixed clone flows for NFG and NFS lie well apart for
$\bar \theta_{13} \geq 1^\circ$. As a general comment, we deduce that the 
combination of the NF golden channel with either the NF silver channel 
or with one of the pair BB/SB could in principle solve these three ambiguities. 

This is not the case for the octant clones. Although a significant reduction 
of the degeneracy can be attained combining the NFG with the NFS or with one of BB/SB, 
it seems that a complete elimination of the octant clones can be achieved only by combining 
the three facilities, NFG, NFS and BB/SB. Nevertheless, even using the combination of all 
the available facilities is quite possible that the experimental uncertainties will not permit 
to solve completely the octant ambiguity for small values of $\bar \theta_{13}$. The ultimate 
sensitivity of the future neutrino experiments to $\delta$ is therefore strongly dependent 
on the success or failure in solving it. To decouple the pair BB/SB (e.g. using the 
Minos OA proposal \cite{MinosOA} for a SuperBeam) can probably be of great help in solving the octant clone.

%%%%%%%%%%%%%%%%%%%%%%%%%%%%%%%%%%%%%%%%%%%%%%%%%%%%%%%%%%%
\section*{Acknowledgments}

We thank B.~Gavela, J. Gomez-Cadenas, P.~Hernandez and P. Migliozzi for useful discussions. 
D.M. acknowledges the European Community's Human Potential Programme under contract 
HPRN-CT-2000-00149 Physics at Colliders for finacial supports.

%%%%%%%%%%%%%%%%%%%%%%%%%%%%%%%%%%%%%%%%%%%%%%%%%%%%%%%%%%%

\appendix

\section{The Analytic Clone Location in Matter} 
\label{appe}

In this appendix we generalize the treatment given for the intrinsic clone
in Sect. \ref{sec:intriclone} to a generic case where the discrete variables $s_{atm}$ and $s_{oct}$
can assume different values between left and right hand side of eqs.~(\ref{eq:ene0})-(\ref{eq:ene0t23sign}).

Using eq.~(\ref{eq:genefunctions}) and eq.~(\ref{eq:gpm}) we get:
\bea
\sin \delta &=& \frac{C^\prime_1 \sin^2 2\bar \theta_{13} - C_1 \sin^2 2 \theta_{13}}{\cos \theta_{13} \sin 2 \theta_{13}}
            + \left [ C^\prime_2 \cos \bar \delta + D^\prime_3 \sin \bar \delta \right ] g (\theta_{13}, \bar \theta_{13})
                    + C^\prime_4 h^\prime (\theta_{13} ) \nn \\ 
& & \label{eq:defcdsuper} \\
\cos \delta &=& \frac{D^\prime_1 \sin^2 2\bar \theta_{13} - D_1 \sin^2 2 \theta_{13}}{\cos \theta_{13} \sin 2 \theta_{13}}
            + \left [ D^\prime_2 \cos \bar \delta + C^\prime_3 \sin \bar \delta \right ] g (\theta_{13}, \bar \theta_{13})
                    + D^\prime_4 h^\prime (\theta_{13} ) \nn
\eea
where, using a notation as much resembling as possible that of the intrinsic case, we have defined
the following coefficients: 
\be
\begin{array}{lll}
\left \{ 
\begin{array}{lll}
     C_1 & = & \dd \left [ \frac{     I^+_1 I^-_2 - I^-_1 I^+_2}{
                                      I^+_3 I^-_2 + I^-_3 I^+_2} \right ]  \\
&& \\
C^\prime_1 & = & \dd \left [ \frac{\bar I^+_1 I^-_2 - \bar I^-_1 I^+_2}{
                                      I^+_3 I^-_2 + I^-_3 I^+_2} \right ]  \\
&& \\
C^\prime_2 & = & \dd \left [ \frac{\bar I^+_2 I^-_2 - \bar I^-_2 I^+_2}{
                                      I^+_3 I^-_2 + I^-_3 I^+_2} \right ]  \\
&& \\
C^\prime_3 & = & \dd \left [ \frac{\bar I^+_3 I^-_3 - \bar I^-_3 I^+_3}{
                                      I^+_3 I^-_2 + I^-_3 I^+_2} \right ]  \\
&& \\
C^\prime_4 & = & \dd \left [ \frac{\bar I_4 (I^-_2 - I^+_2)}{
                                      I^+_3 I^-_2 + I^-_3 I^+_2} \right ]  \\
\end{array}
\right . 
& \qquad &
\left \{ 
\begin{array}{lll}
D_1 & = & \dd \left [ \frac{I^+_1 I^-_3 + I^-_1 I^+_3}{
                            I^+_3 I^-_2 + I^-_3 I^+_2} \right ] \\
&& \\
D^\prime_1 & = & \dd \left [ \frac{\bar I^+_1 I^-_3 + \bar I^-_1 I^+_3}{
                                      I^+_3 I^-_2 + I^-_3 I^+_2} \right ]  \\
&& \\
D^\prime_2 & = & \dd \left [ \frac{\bar I^+_2 I^-_3 + \bar I^-_2 I^+_3}{
                                      I^+_3 I^-_2 + I^-_3 I^+_2} \right ]  \\
&& \\
D^\prime_3 & = & \dd \left [ \frac{\bar I^+_3 I^-_2 + \bar I^-_3 I^+_2}{
                                      I^+_3 I^-_2 + I^-_3 I^+_2} \right ]  \\
&& \\
D^\prime_4 & = & \dd \left [ \frac{\bar I_4 (I^-_3 + I^+_3)}{
                                      I^+_3 I^-_2 + I^-_3 I^+_2} \right ] 
\end{array}
\right .
\end{array}
\label{eq:ciedi}
\ee

Notice that in vacuum all the coefficients of the $C$-type vanish. 
Using the discrete transformation properties of the integrals $I_j$ of 
Tab. \ref{tabella} we derive the corresponding transformation properties for the 
$C_i$ and $D_i$ coefficients, reported in Tab.~\ref{tabella2}. 
The $\theta_{23}$-octant transformation does not affect any of the $C^\prime_i$ or $D^\prime_i$ coefficients, 
but only $C_1$ and $D_1$ that are multiplied by $\cot^2 \theta_{23}$ when $\theta_{23} \to \pi/2 - \theta_{23}$.
The effect of an $s_{atm}$ change is much stronger: all coefficients are affected by this change. 
The $\bar C^\prime_i$ and $\bar D^\prime_i$ coefficients are obtained by the $C_{1,2,3}^\prime,D_{1,2,3}^\prime$ 
in eq.~(\ref{eq:ciedi}) by changing $I^\pm_2 \to - \bar I^\mp_2$ and $I^\pm_3 \to \bar I^\mp_3$.

\begin{table}[h!]
\centering
\begin{tabular}{|c|c|c|c|c|}
\hline
&  {\it Intrinsic} & {\it Octant} & {\it Sign} & {\it Mixed} \\
\hline
& & & & \\
     $C_1$ & $\bar C_1$ & $\cot^2 \theta_{23} \, \bar C_1$ &      $ - \bar C_1$ &  $- \cot^2 \theta_{23} \, \bar C_1$ \\
$C^\prime_1$ & $\bar C_1$ &                       $\bar C_1$ & $\bar C^\prime_1$ &                    $\bar C^\prime_1$ \\
$C^\prime_2$ &        $0$ &                              $0$ & $\bar C^\prime_2$ &                    $\bar C^\prime_2$ \\
$C^\prime_3$ &        $0$ &                              $0$ & $\bar C^\prime_3$ &                    $\bar C^\prime_3$ \\
$C^\prime_4$ & $\bar C_4$ &                       $\bar C_4$ &      $ - \bar C_4$ &                       $ - \bar C_4$ \\
\hline
& & & & \\
     $D_1$ & $\bar D_1$ & $\cot^2(\theta_{23})\, \bar D_1$ &     $ - \bar D_1$ &  $- \cot^2(\theta_{23}) \, \bar D_1$ \\
$D^\prime_1$ & $\bar D_1$ &                       $\bar D_1$ & $\bar D^\prime_1$ &                    $\bar D^\prime_1$ \\
$D^\prime_2$ &        $1$ &                              $1$ & $\bar D^\prime_2$ &                    $\bar D^\prime_2$ \\
$D^\prime_3$ &        $1$ &                              $1$ & $\bar D^\prime_3$ &                    $\bar D^\prime_3$ \\
$D^\prime_4$ & $\bar D_4$ &                       $\bar D_4$ &      $ - \bar D_4$ &                       $ - \bar D_4$ \\
\hline
\end{tabular}
\caption{\it Transformation laws of the coefficients $C_1, C^\prime_i$ and $D_1, D^\prime_i$
under the discrete ambiguities in the $\theta_{23}$ octant and in the $\Delta m^2_{23}$ sign.} 
\label{tabella2}
\end{table}

We present hereafter the explicit analytic solution for the $\theta_{13}$-shift for the octant and sign clones. 
Inserting these results in one of eqs.~(\ref{eq:defcdsuper}) the $\delta$-shift can be extracted.
We do not present here the corresponding expression for the mixed clone. 

\subsection{The Octant Clone}

Using Tab.~\ref{tabella2}, we get for the $\theta_{13}$-shift for the octant clone the following expression: 
\bea
\sin^2 2 \theta_{13}  &=&  \sin^2 2 \bar \theta_{13} + \label{eq:octamatter}\\
& & \nonumber \\
&& \hskip-2cm + \left \{ 
       \frac{\tan^2 \theta_{23}}{\bar C_1^2 + \bar D_1^2} \, \left [ 
         \left ( \bar C_1 \sin \bar \delta + \bar D_1 \cos \bar \delta \right ) \sin 2 \bar \theta_{13} 
                                 + \frac{1}{2} \, \tan^2 \theta_{23} 
                                               \right ] \right . \nonumber \\
&& \hskip-2cm - \left . \left ( 1 - \tan^2 \theta_{23} \right ) \left ( \sin^2 2 \bar \theta_{13} 
- \tan^2 \theta_{23} \, \frac{\bar C_1 \bar C_4 + \bar D_1 \bar D_4}{\bar C_1^2 + \bar D_1^2} \right )
    \right \} \nonumber \\
& & \nonumber \\
&& \hskip-2cm \pm \frac{\tan^2 \theta_{23}}{\sqrt{\bar C_1^2 + \bar D_1^2}} \, \left \{ 
\frac{1}{\bar C_1^2 + \bar D_1^2}
    \left[ \left ( \bar C_1 \sin \bar \delta + \bar D_1 \cos \bar \delta \right ) \sin 2 \bar \theta_{13} 
                                 + \frac{1}{2} \, \tan^2 \theta_{23} \right]^2 \right . \nonumber \\
&& \hskip-2cm - \left . \left ( 1 - \tan^2 \theta_{23} \right ) \left [ \sin^2 2 \bar \theta_{13} - 
\tan^2 \theta_{23} \, \frac{\bar C_1 \bar C_4 + \bar D_1 \bar D_4}{\bar C_1^2 + \bar D_1^2} \right ] \right . \nn \\
&& \hskip-2cm - \left . 2 ( 1 - \tan^2 \theta_{23} ) \left [ 
\left ( \bar C_4 \sin \bar \delta + \bar D_4 \cos \bar \delta \right ) - 
\frac{ (\bar C_1 \bar C_4 + \bar D_1 \bar D_4) (\bar C_1 \sin \bar \delta + \bar D_1 \cos \bar \delta)}{
 \bar C_1^2 +\bar D_1^2} \right ] \sin 2 \bar \theta_{13} \right . \nn \\
&& \hskip-2cm - \left . ( 1 - \tan^2 \theta_{23} )^2 \left [ (\bar C_4^2 + \bar D_4^2) 
-  \frac{(\bar C_1 \bar C_4 + \bar D_1 \bar D_4)^2}{\bar C_1^2 + \bar D_1^2}
\right ]
                                          \right \}^{1/2} \nn
\eea

It can be easily shown that in the vacuum limit, when $C_1, C^\prime_i \to 0$, 
this equation reduces to the vacuum solution, eq.~(\ref{eq:octavacuum2}). Moreover, for $\theta_{23} \to 45^\circ$ reduces to 
the general solution in matter, eq.~(\ref{eq:solEP}), for the intrinsic clone. 

\subsection{The Sign Clone}

Using Tab.~\ref{tabella2}, we get for the $\theta_{13}$-shift for the sign clone the following expression: 
\bea
\label{eq:signmatter}
\sin^2 2 \theta_{13} & = & - \left ( \frac{\bar C_1 \bar C_1^\prime + \bar D_1 \bar D_1^\prime}{\bar C_1^2 + \bar D_1^2} \right ) 
\sin^2 2 \bar \theta_{13}
                        \\
&& \hskip-2cm + \frac{1}{2 \left ( \bar C_1^2 + \bar D_1^2 \right ) } \, 
    \left [ 1 - 2 \left ( \bar C_1 \bar C_2^\prime + \bar D_1 \bar D_2^\prime \right ) \cos \bar \delta \sin 2 \bar \theta_{13} 
              - 2 \left ( \bar C_1 \bar D_3^\prime + \bar D_1 \bar C_3^\prime \right ) \sin \bar \delta \sin 2 \bar \theta_{13}
    \right ] \nn \\
&& \hskip-2cm \pm \frac{1}{2 \left ( \bar C_1^2 + \bar D_1^2 \right ) } \, 
        \left \{
\left [ 1 - 4 \left ( \bar C_1 \bar C_2^\prime + \bar D_1 \bar D_2^\prime \right ) \cos \bar \delta \sin 2 \bar \theta_{13} 
          + 4 \left ( \bar C_1 \bar C_3^\prime + \bar D_1 \bar D_3^\prime \right )^2 \cos^2 \bar \delta \sin^2 2 \bar \theta_{13}
\right ] \right . \nn \\
&& \hskip-2cm - 4 \left [ \left ( \bar C_1 \bar C_3^\prime + \bar D_1 \bar D_3^\prime \right )^2 
             + \left ( \bar C_1 \bar C_1^\prime + \bar D_1 \bar D_1^\prime \right ) \right ]  \sin^2 2 \bar \theta_{13} \nn \\
&& \hskip-2cm - 4 \left ( \bar C_1 \bar D_1^\prime - \bar D_1 \bar C_1^\prime \right )^2 \sin^4 2 \bar \theta_{13} \nn \\
&& \hskip-2cm - 8 \left ( \bar C_1 \bar D_1^\prime - \bar D_1 \bar C_1^\prime \right ) 
      \left [ \left ( \bar C_1 \bar D_2^\prime - \bar D_1 \bar C_2^\prime \right ) \cos \bar \delta + 
              \left ( \bar C_1 \bar C_3^\prime - \bar D_1 \bar D_3^\prime \right ) \sin \bar \delta     
     \right ] \sin^3 2 \bar \theta_{13} \nn \\
&& \hskip-2cm - 4 \left ( \bar C_1 \bar D_2^\prime + \bar D_1 \bar C_2^\prime \right )^2 \cos^2 \bar \delta \sin^2 2 \bar \theta_{13} \nn \\
&& \hskip-2cm - 8 \left ( \bar C_1 \bar D_2^\prime - \bar D_1 \bar C_2^\prime \right ) 
       \left ( \bar C_1 \bar C_3^\prime - \bar D_1 \bar D_3^\prime \right ) 
                               \sin \bar \delta \cos \bar \delta \sin^2 2 \bar \theta_{13} \nn \\
&& \hskip-2cm \left . 
   - 4 \left ( \bar C_1 \bar D_3^\prime + \bar D_1 \bar C_3^\prime \right ) \sin \bar \delta \sin 2 \bar \theta_{13}
              \right \}^{1/2} \nn
\eea

One can verify that eq.~(\ref{eq:signmatter}) reduces to the vacuum solution, eq.~(\ref{eq:thetavacuum2}). 
The first term in the square root goes to a perfect square; the second vanishes for an exact cancellation between 
$( \bar D_1 \bar D_3^\prime )^2$ and $\bar D_1 \bar D_1^\prime$, since in vacuum 
$\bar D_1^\prime \to - \bar D_1, \bar D_2^\prime \to -1$ and $\bar D_3^\prime \to 1$. 
All other terms in the square root trivially vanish since are multiplied by some $C$-type coefficient that goes to zero 
in the vacuum limit.

%%%%%%%%%%%%%%%%%%%%%%%%%%%%%%%%%%%%%%%%%%%%%%%%%%%%%%%%%%%

%%%%%%%%%%%%%%%%%%%%%%%%%%%%%%%%%%%%%%%%%%%%%%%%%%%%%%%%%%%
\end{document}